\documentclass[11pt,a4paper]{article}
\pdfoutput=1
\usepackage{jheppub}
\usepackage[T1]{fontenc}
\usepackage[utf8]{inputenc}

\usepackage{subcaption}
\usepackage{amsmath}
\usepackage{xspace}
\usepackage{mathrsfs}
\usepackage{physics}
\usepackage{orcidlink}

\usepackage{cleveref}

\numberwithin{equation}{section}
\newcommand{\nn}{\nonumber}

\newcommand{\bOm}{\overline{\Omega}}
\newcommand{\cE}{\mathcal{E}}
\newcommand{\cV}{\mathcal{V}}
\newcommand{\cQ}{\mathcal{Q}}

\newcommand{\cB}{\ensuremath{\mathcal B}}
\newcommand{\cD}{\ensuremath{\mathcal D}}

\newcommand{\cI}{\ensuremath{\mathcal I}}

\newcommand{\cM}{\ensuremath{\mathcal M}}
\newcommand{\cN}{\ensuremath{\mathcal N}}

\newcommand{\IR}{\ensuremath{\mathbb R}}
\newcommand{\IZ}{\ensuremath{\mathbb Z}}

\newcommand{\IP}{\ensuremath{\mathbb P}}
\newcommand{\IQ}{\ensuremath{\mathbb Q}}
\newcommand{\IT}{\ensuremath{\mathbb T}}

\newcommand{\I}{{\mathrm i}}

\newcommand{\de}{\mathrm{d}}

\newcommand{\Ux}{\mathsf{U}_x}
\newcommand{\Uy}{\mathsf{U}_y}
\newcommand{\Uz}{\mathsf{U}_z}

\DeclareMathOperator{\Erf}{Erf}
\DeclareMathOperator{\Erfc}{Erfc}
\DeclareMathOperator{\sign}{sign}
\DeclareMathOperator{\sgn}{sign}
\DeclareMathOperator{\ch}{ch}
\DeclareMathOperator{\Todd}{Todd}
\DeclareMathOperator{\Sym}{Sym}
\DeclareMathOperator{\Vol}{Vol}
\DeclareMathOperator{\Ind}{Ind}

\newcommand{\sfE}{\mathsf{E}}
\newcommand{\sfM}{\mathsf{M}}
\newcommand{\sfH}{\mathsf{H}}
\newcommand{\sfL}{\mathsf{L}}

\def\bea{\begin{eqnarray}}
\def\eea{\end{eqnarray}}
\def\be{\begin{equation}}
\def\ee{\end{equation}}
\def\ba{\begin{align}}
\def\ea{\end{align}}
\def\bse{\begin{subequations}}
\def\ese{\end{subequations}}

\newcommand{\ie}{i.e.\@\xspace}

\title{Black Hole Quantum Mechanics \\ and Generalized Error Functions}

\preprint{arXiv:2507.08551
}

\author{Boris Pioline \orcidlink{0000-0002-3168-288X}}
\author{and Rishi Raj \orcidlink{0000-0002-1310-1025}}
\affiliation{Laboratoire de Physique Th\'eorique et Hautes
Energies, CNRS and Sorbonne Universit\'e, \\ 
Campus Pierre et Marie Curie,
4 place Jussieu, F-75252 Paris cedex 05, France}
\emailAdd{pioline@lpthe.jussieu.fr}
\emailAdd{raj@lpthe.jussieu.fr}

\abstract{In Type II Calabi-Yau string compactifications, S-duality predicts that suitable generating
series of BPS indices counting microstates of D4-D2-D0 black holes are in general mock modular forms of higher depth. The non-holomorphic contributions needed to cancel the anomaly under modular transformations involve certain indefinite theta series with kernels constructed from generalized error functions. Physically, these contributions are expected to arise from a spectral asymmetry in the continuum of scattering states of $n$ BPS dyons with mutually non-local charges.
For $n=2$,  the (standard, depth one) error function completion was derived long ago by explicitly computing the bosonic and fermionic density of states in the two-body supersymmetric quantum mechanics. Here we derive the general non-holomorphic completion for an arbitrary number of centers by evaluating the refined Witten index of the supersymmetric quantum mechanics
using localization. In a nutshell, the index reduces to an integral over $\IR^{3n-3}$ (the relative location of the centers), and splits into an integral over the $2n-2$ dimensional phase space of BPS ground states times an integral over $n-1$ transverse directions, which ultimately produces the expected generalized error functions.
}

\begin{document}

\maketitle
\flushbottom

\section{Introduction}

In string vacua with extended supersymmetry, an important challenge is to count BPS states with given electromagnetic charge $\gamma$, and match its  logarithm with the Bekenstein-Hawking entropy of the corresponding supersymmetric black hole in the large charge regime. While this is usually a formidable task, it becomes feasible when suitable generating series of BPS indices possess modular properties, since the asymptotic growth of Fourier coefficients of modular forms is then under excellent analytic control. Such modular properties are guaranteed when the BPS black holes arise as BPS black strings wound on a circle, as the generating series of indices can be identified with the elliptic genus of a superconformal field theory (SCFT) on the string worldsheet~\cite{Strominger:1996sh}. 

This applies in particular to D4-D2-D0 black holes in Type IIA strings compactified on a Calabi-Yau (CY) threefold, as they arise from the M-theory viewpoint as M5-branes wrapped on a divisor $\cD$ determined by the D4-charge, leading to a black string in 5 dimensions, further wrapped on the eleven-dimensional circle. If the divisor $\cD$ is very ample and irreducible, the black string SCFT has a discrete spectrum (after factoring out the center of mass degrees of freedom) and the growth of the Fourier coefficients of the elliptic genus precisely matches the Bekenstein-Hawking-Wald prediction~\cite{Maldacena:1997de}.
In general, however, if the divisor class $[\cD]$ is reducible into a sum $\sum_{i=1}^n [\cD_i]$ of irreducible classes, the black string can split into $n$ constituents, depending on the values of the moduli at infinity. The elliptic genus in a suitable chamber (or rather, the vector-valued modular form arising from its theta series decomposition) then  becomes mock modular of depth $n-1$, with a specific modular anomaly which has been characterized in a series of recent works \cite{Alexandrov:2016tnf,Alexandrov:2017qhn,Alexandrov:2018lgp,Alexandrov:2019rth} (see \cite{Alexandrov:2025sig} for a review).

This modular anomaly (or mock modular behavior) was first observed in the context of $\cN=4$ string vacua, such as Type II strings on $K3\times T^2$ or heterotic strings on $T^6$, where the only possible splittings involve only $n=2$ constituents, and the BPS indices in the attractor chamber are given by
Fourier coefficients of mock modular forms of depth one \cite{Dabholkar:2012nd}, similar to Ramanujan's famous mock theta series. As discovered in~\cite{Zwegers-thesis} (see also~\cite{MR2605321} for a broader historical account), their non-holomorphic completion is given by a certain indefinite theta series of signature $(1,1)$, with a kernel expressed as a linear combination of error functions of the lattice vector coordinates, decaying exponentially outside the positive light-cone. The fact that the error function satisfies a specific, second-order partial differential equation known as Vign\'eras' equation, along with the
decay property of the kernel, ensures the convergence and modularity of the completed theta 
series~\cite{Vigneras:1977}. 
Physically, the non-holomorphic completion can be traced to the contribution of the spectral asymmetry of the continuum of  scattering states  in the quantum mechanics of two BPS black 
holes~\cite{Pioline:2015wza}, or equivalently in the SCFT of two BPS black strings~\cite{Murthy:2018bzs} (see \cite{Troost:2010ud,Eguchi:2010cb,Ashok:2011cy,Murthy:2013mya,Ashok:2013pya,Harvey:2014nha,Gupta:2017bcp,KumarGupta:2018rac,Gaiotto:2019gef} for other examples of superconformal field theories with continuous spectrum where the elliptic genus turns out to be non-holomorphic). Indeed, the two-body quantum mechanics  is simple enough that the density of bosonic and fermionic states can be computed separately --- their difference leads to a complementary error function  of the Fayet-Iliopoulos (FI) stability parameter (namely, $M_1(x)$ defined in \eqref{eqn:defM1} below), which 
combines with the step function $\sign(x)$ in the bound state contribution to produce a 
smooth error function ($E_1(x)$ defined in \eqref{eqn:defE1} below) across the wall of marginal stability~\cite{Pioline:2015wza}.

For a general CY threefold $X$ with $SU(3)$ holonomy, the number $n$ of bound state constituents can be arbitrarily large, and the modular completion of the generating series of BPS indices in general involves
an indefinite theta series of signature $(n-1,(n-1) (b_2(X)-1))$, whose kernel is expressed as a specific combination of so-called generalized error functions $E_n(\cM,x), M_n(\cM,x)$ of the lattice vector coordinates. These special functions were introduced in \cite{Alexandrov:2016enp} for the purpose of constructing indefinite theta series satisfying Vign\'eras' criteria~\cite{Vigneras:1977} for convergence and modularity. They are natural generalizations of the standard and complementary error functions\footnote{The latter can also be expressed in terms of the incomplete Gamma functions, $E_1(x)=\gamma(\frac12,\pi x^2)/\sqrt{\pi}, M_1(x)=-\sign(x) \Gamma(\frac12,\pi x^2)/\sqrt{\pi}$, but the representations \eqref{eqn:defE1}, \eqref{eqn:defM1} are the ones which generalize most directly.} 
\bea
E_1(x) &=&\int_{\IR}  e^{-\pi (x-x')^2} \sign(x') \, \de x'= \Erf(x\sqrt\pi)  \label{eqn:defE1}, \\
M_1(x) &=&\frac{\I}{\pi} \int_{\IR-\I x} e^{-\pi z^2-2\pi\I z} \frac{\de z}{z} = -\sign(x) \label{eqn:defM1}
\Erfc(|x|\sqrt\pi),
\eea
now involving a $(n-1)$-dimensional Gaussian integral,  a product of $n-1$  factors $\sign(C_i,x)$ in the case of $E_n$, or a product of $n-1$ poles $1/(C_i^\star,z)$ in the case of $M_n$, where the vectors $C_i$ and $C_i^\star$ are determined by the matrix $\cM$. While $E_n(\cM,x)$ is a smooth $C^\infty$ function of $x$, $M_n(\cM,x)$ is discontinuous on hyperplanes $(C_i^\star,x)=0$ but the two can be expressed in terms of each other through analogues of the relation 
$E_1(x)= M_1(x)+\sign(x)$ \cite{Nazaroglu:2016lmr}. Our goal in this note is to show how these non-holomorphic completion terms arise from the Witten index of the supersymmetric quantum mechanics of $n$ BPS dyons. 

More specifically, we shall compute the Witten index $\cI_n=\Tr(-1)^F e^{-\beta H}$ by localization, reducing the functional integral to a finite-dimensional integral over time-independent configurations. We will see that the integral over the relative locations of the $n$ centers in $\IR^3$, along with their fermionic partners, can be rewritten as an integral over the $2n-2$-dimensional phase space $\cM_n(\{\gamma_i,u_i\})$ of supersymmetric multi-centered black holes with charges $\{\gamma_i\}$ for fixed values of the 
(dynamical) FI parameters $\{u_i\}$, times an integral over the $n-1$ transverse directions corresponding to varying the $u_i$'s, with a Gaussian measure factor depending on the (fixed, physical) FI parameters $c_i$. The former integral being locally constant in the $u_i$'s, the latter leads to a linear superposition of generalized error functions, as required for the modular completion.

The remainder of this note is organized as follows. In Section \ref{sec:review}, we review the
prescription for the non-holomorphic modular completion of generating series of D4-D2-D0
indices, and define the generalized error functions from the title. In Section \ref{sec:witten-index}, after recalling basic aspects of the 
supersymmetric quantum mechanics of $n$ dyons, we compute the refined Witten index  using localization (see Section \ref{sec:computing-witten-index} and Appendix \ref{sec:proof-master-identity} for the actual computation). In Section \ref{sec:comparison}, we compare this result to the modular prediction for $n=3$ and $n=4$ centers  in the special case $X=K_{\IP^2}$, and find perfect agreement with the results of \cite{Alexandrov:2019rth}, at least when the fugacity $y$ is a pure phase. We give further evidence that the supersymmetric quantum mechanics reproduces the instanton generating potential for an arbitrary CY threefold.

\section{Modular completions and generalized error functions \label{sec:review}}

In this section we briefly review the mock modular properties of generating series of 
D4-D2-D0 indices in Type IIA strings compactified on a general smooth Calabi-Yau threefold $X$, 
and their non-holomorphic completion in terms of generalized error functions. 

In this context,
BPS states are identified mathematically as semi-stable objects $E$ in the derived category of coherent sheaves
$D^b{\rm Coh}(X)$. They carry the `electromagnetic' charge $\gamma(E)=\ch E \sqrt{\Todd(X)}\in H^{\rm even}(X,\IQ)$,
with components $\gamma=(p^0,p^a,q_a,q_0)$ on a basis $(1,\omega_a,\omega^a,\Vol X)\in
H^0\oplus H^2\oplus H^4\oplus H^6$ corresponding respectively to the D6-, D4-, D2- and D0-brane charges (here 
 $\omega_a, a=1\dots b_2(X)$ denotes a basis of the K\"ahler cone in $H^2(X,\IZ)$, and $\omega^a$ is the 
 Poincar\'e dual basis in $H^4(X,\IZ)$, such that $\omega_a \omega^b = \delta_a^b \Vol X$). The mass of a BPS state of charge $\gamma$ is given (in Planck units) by the absolute value of the central charge $Z(\gamma)$, which in the large volume limit takes the form
 \be
 Z_{\rm LV}(\gamma) = q_0+ q_a z^a +  \frac12 \kappa_{abc} p^a z^b z^c - \frac16 p^0 \kappa_{abc} z^a z^b z^c.
 \ee
Here $z^a=b^a+\I t^a$ parametrize the complexified 
K\"ahler moduli space $\cM_K$ and 
 $\kappa_{abc}=\int_X \omega_a  \omega_b  \omega_c$ is the cubic intersection form.  The Dirac-Schwinger-Zwanziger (DSZ) product is given by the antisymmetrized Euler form $\langle \gamma,\gamma'\rangle = p'^0 q_0 + p'^a q_a - p^a q'_a - p^0 q'_0$. Wall-crossing for a fixed charge $\gamma$ may occur along real codimension-one loci in $\cM_K$ where the central charges $Z(\gamma_i)$ of two  charge vectors $\gamma_1, \gamma_2$ with $\gamma=N_1 \gamma_1+N_2\gamma_2$ become aligned, with $N_1,N_2\geq 1$ and $\langle \gamma_1,\gamma_2\rangle\neq 0$. 
 
When the D6-brane charge $p^0$ 
vanishes (the main case of interest in this note), the central charge 
at leading order $Z_{\rm LV}(\gamma) \simeq -\frac{1}{2} (p t^2) + i (q_a + (p b)_a)t^a$
and the DSZ product $\langle \gamma_1,\gamma_2\rangle$  
depend only on the reduced charge vector   $\check\gamma=(p^a,q_a)$. In the large volume limit, wall-crossing may occur when the following quantity vanishes,
\be
\label{eqn:ImZ1Z2}
\cI_{\gamma_1,\gamma_2} = \frac{ \Im(\bar{Z}_{\rm LV}(\gamma_1) Z_{\rm LV}(\gamma_2)) }
{\sqrt{2 |Z_{\rm LV}(\gamma_1) Z_{\rm LV}(\gamma_2) Z_{\rm LV}(\gamma)|}}
= 
\frac{(p_2 t^2) (q_{1, a} + (b p_1)_a) t^a - (p_1 t^2) (q_{2, a} + (b p_2)_a) t^a}
{\sqrt{(p_1 t^2) (p_2 t^2) (p t^2)} },
\ee
where we use the shorthand notations $(pt^2)=\kappa_{abc} p^a t^b t^c$, $(bp)_a = \kappa_{abc} b^b p^c$. The factors in the denominator ensure that $\cI_{\gamma_1,\gamma_2}$ is invariant under rescaling $t^a\to \lambda t^a$ with $\lambda>0$, and are recognized as products of the masses
$m_i = \frac12 (p_i t^2)$ in the large-volume limit.

\subsection{Mock modularity of D4-D2-D0 invariants} 
\label{sec:mock-modularity}
Let $\bOm(\gamma,z^a,y)$ be the refined, rational Donaldson-Thomas (DT) invariant
counting semi-stable BPS states of charge $\gamma$ with respect to the stability condition
associated to the point $z^a=b^a+\I t^a$. 
$\bOm(\gamma,z^a,y)$ is  in general a rational function of $y$, but it is expected that the
`integer DT invariant'
\be
\Omega(\gamma,z^a,y):= \sum_{d|\gamma} \mu(d) \frac{y-1/y}{y^d-1/y^d} 
\bOm(\gamma/d,z^a,y^d),
\ee
is a symmetric Laurent polynomial in $y$ with integer coefficients, corresponding to the 
 physical refined BPS index (here $\mu(d)$ is the multiplicative M\"obius function). This refined index  in general depends on the complex structure moduli (except for toric threefolds, in which case it is protected),
 but its `unrefined' limit as $y\to 1$ depends only on the K\"ahler moduli (or more generally, on the choice of Bridgeland stability condition). As explained in \cite{Alexandrov:2019rth}, even if one is only interested
in unrefined indices,  it is advantageous to introduce the refinement  
 at intermediate stages, as it allows to express the modular completion in a much more concise way. 
 
 \medskip
 
 We now focus on the so-called MSW invariants, counting objects with vanishing D6-brane charge $p^0$ and fixed D4-brane charge supported on an ample divisor $p^a\in \Lambda^*=H^2(X,\IZ)$ in the `large volume attractor chamber'
 $t^a= \lambda p^a, b^a=-\kappa^{ab} q_b$ where $\kappa^{ab}$ is the inverse of the quadratic form $\kappa_{ab}:=\kappa_{abc} p^c$ and $\lambda$ is taken arbitrarily large and positive. By spectral flow invariance (\ie tensoring by a flat line bundle), the index $\bOm(\gamma,z^a,y)$ 
only depends on the `invariant D0-brane charge' 
 $\hat q_0\in \IQ$ and `discrete flux' $\mu_a\in \Lambda^*/\Lambda$ defined by 
\be
\hat q_0 = q_0 -\frac12 \kappa^{ab} q_a q_b, \quad 
q_a = \mu_a + \frac12 \kappa_{ab} p^b  + \kappa_{ab} \epsilon^b,
\ee
where $\epsilon^a\in \IZ$ are the spectral flow parameters. 
We define the MSW invariants as 
\be
\bOm_{p^a,\mu^a}(\hat q_0) := \lim_{\lambda\to+\infty} \bOm((0,p^a,q_a,q_0), -\kappa^{ab} q_b
+  \I \lambda p^a,y).
\ee
Since $\hat q_0$ is bounded from above by the Bogomolov bound, we can define the generating series
\be
h_{p,\mu}(\tau,y) =\frac{1}{y-1/y} \sum_{\hat q_0\leq \hat q_0^{\rm max}} 
\bOm_{p^a,\mu^a}(\hat q_0)\, e^{-2\pi\I \hat q_0 \tau},
\ee
where $\tau$ lives in the Poincar\'e upper half-plane, $\tau_2:=\Im\tau>0$. 
The S-duality symmetry of string theory (which is manifest from the viewpoint of M-theory compactified on $X\times T^2$)
then implies that the generating series $h_{p,\mu}(\tau,y), \mu\in\Lambda^*/\Lambda$ should in general transform under 
$(\tau,w)\mapsto (\frac{a\tau+b}{c\tau+d},\frac{w}{c\tau+d})$ (with $y:=e^{2\pi\I w}$) as a vector-valued mock Jacobi form\footnote{To be precise, as currently understood, S-duality only implies that the unrefined limit 
$h_{p,\mu}(\tau):=\lim_{y\to 1} (y-1/y) h_{p,\mu}(\tau,y)$ transforms
as a vector-valued mock modular form of depth $n-1$. For non-compact CY threefolds of the form
$X=K_S$ where $S$ is a Fano surface, such that MSW invariants supported on $r[S]$ coincide
with rank $r$ Vafa-Witten invariants \cite{Vafa:1994tf}, 
it is expected that the generating series of refined VW invariants
are mock Jacobi form of depth $r-1$ \cite{gottsche2018refined,Thomas:2018lvm}, and it was checked in \cite{Alexandrov:2019rth} that the prescription below for constructing the modular completion reproduces known results for $S=\IP^2$ and rank $r\leq 3$ \cite{Bringmann:2010sd,Manschot:2017xcr}, and further used in \cite{Alexandrov:2020bwg,Alexandrov:2020dyy} to predict VW invariants for arbitrary del Pezzo surfaces and rank. Conservatively, the modularity conjecture only applies to the unrefined limit.}
of depth $n-1$, where
$n$ is the maximal number of irreducible constituents in the divisor class $p$. This means that
they admit a non-holomorphic completion of the form
\be
\label{eqn:nhcompJ}
\widehat{h}_{p,\mu}(\tau,y) := h_{p,\mu}(\tau,y) +  \sum_{n\geq 2} 
\sum_{\substack{p=\sum_{i=1}^n p_i \\
\mu=\sum_{i=1}^n \mu_i \mod \Lambda}}
\Theta_\mu( \{p_i,\mu_i\}, \tau,y)\, \prod_{i=1}^n
h_{p_i,\mu_i}(\tau,y)\ ,
\ee
where $\Theta_\mu( \{p_i,\mu_i\}, \tau,y) $ are universal indefinite theta series 
such that $\widehat{h}_{p,\mu}(\tau,y)$ transforms as a (ordinary, but non-holomorphic) vector-valued Jacobi form of fixed weight, 
 index and multiplier system. The  indefinite theta series 
are neither holomorphic in $w$ nor modular covariant, but they are constructed such that their modular anomaly cancels that of $h_{p,\mu}$. Moreover, $\Theta_\mu(\{p_i,\mu_i\}, \tau,y)$ has a zero of order $n-1$ at $y=1$, such that \eqref{eqn:nhcompJ} reduces in the unrefined limit to
\be
\widehat{h}_{p,\mu}(\tau) := h_{p,\mu}(\tau) +  \sum_{n\geq 2} 
\sum_{\substack{p=\sum_{i=1}^n p_i \\ \mu=\sum_{i=1}^n \mu_i \mod \Lambda }}
\Theta_\mu( \{p_i,\mu_i\}, \tau)\, \prod_{i=1}^n
h_{p_i,\mu_i}(\tau)\ ,
\ee
with $\Theta_\mu( \{p_i,\mu_i\}, \tau)=\lim_{y\to 1} (y-1/y)^{1-n}\Theta_\mu( \{p_i,\mu_i\}, \tau,y)$. 
The theta series, $\Theta_\mu( \{p_i,\mu_i\}, \tau,y)$, are considerably simpler than their
unrefined counterparts, and given by an indefinite theta series
\be
\label{eqn:defThmu}
\Theta_\mu( \{p_i,\mu_i\}, \tau,y)
=
\sum_{\substack{ q_{a,i} \in \Lambda+\mu_a+\frac12 \kappa_{ab,i} p^b_i \\
\sum_{i=1}^n q_{a,i}=\mu_a+\frac12 \kappa_{ab} p^b}}
R_n\left(\{ p_i, q_i\}, \tau_2,y\right)\, e^{\I \pi\tau Q_n\left(\{ p_i, q_i\}\right)},
\ee
where $\kappa_{ab,i}=\kappa_{abc} p_i^c$, $\kappa_i^{ab}$ is its inverse and $Q_n$ is the
quadratic form
\be
Q_n\left(\{ p_i, q_i\}\right):= \kappa^{ab} q_a q_b  - \sum_{i=1}^n \kappa_i^{ab} q_{i,a} q_{i,b},
\ee
of signature $(n-1)(b_2(X)-1,1)$ on the lattice specified in the summation range in \eqref{eqn:defThmu}. 
The kernel $R_n\left(\{ p_i, q_i\}, \tau_2,y\right)$ is given by 
a sum of generalized error functions multiplied by step functions of the $q_i$'s, determined as follows:\footnote{We follow the conventions in \cite{Alexandrov:2024jnu,Alexandrov:2025sig}. The symbol $\Sym$ indicates that one should sum over all
 permutations of $\{(p_i,q_i)\}$ with weight $1/n!$.}
\be
R_n(\{p_i,q_i\},\tau_2,y)= \Sym\left\{
\frac{(-y)^{\sum_{i<j} \langle \alpha_i, \alpha_j \rangle} }{2^{n-1}} 
\sum_{T\in\IT_n^{\rm S}}(-1)^{|V_T|-1}
(  \cE_{v_0} - \cE_{v_0}^{(0)} )
 \prod_{v\in V_T\setminus{\{v_0\}}}  \cE_v^{(0)}\right\}.
\label{eqn:solRnr}
\ee
Here, $T$ runs over all rooted planar trees  and 
valency $\geq 3$ (also called Schr\"oder trees) with $n$ leaves; 
$V_T$ denotes the set of vertices (excluding the leaves) and $v_0$ the root vertex;
one then associates a reduced charge vector $\check\gamma_v=(p_v,q_v)$ to each vertex such that the 
leaves carry charges  $\check\gamma_i=(p_i^a,q_{a,i})$ 
and the charges add up at each vertex. For each vertex,
$\cE_v^{(0)}$ is a product of $m-1$ sign factors ($m$ being the number of descendants
of the vertex $v$, and $\{\alpha_i\}$  being the corresponding reduced charges)\footnote{There
is a subtlety in interpreting the formula in \eqref{eqn:defEv0} when some of the  $\Gamma_k$'s vanish. 
If the  subset $\cI\subset \{1,\dots,m-1\}$ such that $\Gamma_k=0$
has even cardinality, then the product 
$\prod_{k=1}^{m-1} \sign\Gamma_k$ should be replaced by 
$\frac{1}{|\cI|+1} \prod_{k\notin\cI} \sign\Gamma_k$. If the set $\cI$ has odd cardinality, then $\cE_v^{(0)}$ should be taken to vanish.}
\be
\label{eqn:defEv0}
\cE_v^{(0)} = 
\prod_{k=1}^{m-1} 
\sign\Gamma_k, \quad \Gamma_k:=
\left( \sum_{i=1}^k \sum_{j=k+1}^m \langle \alpha_i, \alpha_j \rangle \right).
\ee
Finally, the factor $\cE_{v_0}$ attached to the root vertex is a depth $m-1$ generalized boosted error function (defined in the next subsection, \cref{eqn:defEgen})
\be\label{eqn:def-Ev0-erf}
\cE_{v_0} = 
\, \Phi^{E}_{m-1}(\cQ,\{ \vb*{v}_k\},   \vb{x} ).
\ee
Here bold letters denote elements of a vector space which is $n$ copies of $\mathbb{R}^{1, b_2(X)-1}$ (a vector $\vb*{x}$ has components $x^a_i$) endowed with a quadratic form $\cQ$ given by $\vb*{x} \, \cdot \, \vb*{y} = \sum_{i=1}^{n} (p_i x_i y_i)$. The vectors $\{ \vb*{v}_k, k=1,\dots m\}$ in \eqref{eqn:def-Ev0-erf} are 
chosen such that 
\be
\label{eqn:vdotx}
\vb*{v}_{k} \cdot \vb*{x} = \sum_{i=1}^k \sum_{j=k+1}^m \kappa_{abc} p_i^a p_j^b (x_i^c-x_j^c).
\ee
The vector $\vb*{x}=(x_1^a,\dots, x_n^a)$ is in turn related to the electric charges $q_{a,i}$ by
\be
x_i^a =\sqrt{2\tau_2} \left( \kappa_i^{ab} q_{b,i} + \eta \rho_i^a \right),
\ee
where $\eta:=\frac{\Re(\log y)}{2\pi\tau_2}$ and $\rho_i^a:=-\sum_{j<i} p_j^a + \sum_{j>i} p_j^a$.
The $\eta$-dependent shift appears whenever the refinement parameter $y=e^{2\pi\I w}$ is not a pure phase, and is necessary to resolve ambiguities in defining $\sign(x)$ for $x=0$.  Note  that in the limit $\tau_2\to + \infty$ (after setting $\eta=0$), $\cE_{v_0}$ reduces to the same product of sign functions
as $\cE_{v_0}^{(0)}$, so the factor  $(  \cE_{v_0} - \cE_{v_0}^{(0)} )$ associated to the root vertex 
in \eqref{eqn:solRnr} is actually a generalized complementary error function, denoted by $\cE^{{\rm ref}(+)}_{v_0}$ in \cite[(3.13)]{Alexandrov:2019rth}.  

\medskip

Finally, we mention that the completed generating series $\widehat{h}_{p,\mu}$ enter in the `instanton generating potential' $\mathcal{G}$, whose refined version was introduced in \cite[Section  3.3]{Alexandrov:2019rth},  
\be
\label{eqn:defG}
\mathcal{G} = \sum_{n=1}^{\infty} \frac{1}{2\pi \sqrt{\tau_2}}
\, e^{-\sum_i S_{p_i}^{\rm cl}} 
\vartheta_{p,\mu}( \Phi^{\rm tot}_n,-1)
\left[ \prod_{i=1}^n \sum_{p_i,\mu_i}\sigma_{p_i} h_{p_i,\mu_i}
\right] .
\ee
Here $S_p^{\rm cl}=\pi\tau_2 (pt^2)-2\pi\I p^a \tilde c_a$ is the `classical action', $\tilde c_a$ being a chemical potential conjugate to the magnetic charge $p^a$, 
$\sigma_{p_i}$ are `quadratic refinement' factors defined in 
\cite[(D.5)]{Alexandrov:2018lgp}, and we have set $\eta=0$ for simplicity. 
The kernel $\Phi_n^{\rm tot}$ in the indefinite theta series is again obtained as a sum over Schr\"oder trees \cite[(C.17)]{Alexandrov:2019rth}
\be
\label{eqn:Phitot}
\Phi_n^{\rm tot} = \frac{(-y)^{\sum_{i<j} \langle \alpha_i, \alpha_j \rangle} }{2^{n-1}} \Phi_1^{\int} \sum_{T\in\IT_n^{\rm S}}(-1)^{|V_T|-1}
 (  \widetilde{\cE}_{v_0} - \cE^{(0)}_{v_0} )
\prod_{v\in V_T\setminus{\{v_0\}}}  \cE^{(0)}_v,
\ee
where $\Phi_1^{\int}$ is a Gaussian factor ensuring convergence for the total D2-brane charge $x=\sum \vb{x}_i$, 
\be
\label{eqn:Phi1}
\Phi_1^{\int}(x)=\frac{e^{-\frac{\pi (pxt)^2}{(pt^2)}}}{2\pi\sqrt{2\tau_2 (pt^2)}},
\ee
and $\widetilde\Phi_v$ is given for a vertex with $m$ descendants by 
\be
\widetilde{\cE}_{v}=
 \Phi^{E}_{m-1}(\cQ,\{ \vb*{u}_k\},   \vb{x} ).
\ee
Here, $\vb*{u}_k, k=1\dots m-1$  are moduli-dependent vectors chosen such that (compare with \eqref{eqn:vdotx})
\be
\label{eqn:udotx}
\vb*{u}_{k} \cdot \vb*{x} = \sum_{i=1}^{k}
\sum_{j=k+1}^m  \left[ (p_j t^2) (p_i x_i t) - (p_i t^2) (p_j x_j t) \right].
\ee
We note that $\mathcal{G}$ can be just as well represented as a sum over monomials in the 
completed generating series $\widehat{h}_{p_i,\mu_i}$, at the expense of replacing 
the locally constant $\cE^{(0)}_{v}$ in \eqref{eqn:Phitot} by the full error function $\cE_{v}$ \cite[(3.31)]{Alexandrov:2019rth}. 
These definitions ensure that $\mathcal{G}$ is a completely smooth function of the moduli (away from conifold singularities),   transforming as a Jacobi form of 
weight $(-\frac12,\frac12)$ under $SL(2,\IZ)$. Its residue at $y=1$ is closely related to the contact potential on the moduli space after compactification on a circle, and plays the role of the Witten index for the full 4D supergravity theory on $\IR^3\times S^1$ \cite{Alexandrov:2014wca}. 

\medskip

While the prescriptions summarized above appear to be complicated, it is straightforward to implement them on a symbolic computation system. For $n=2$, the theta series \eqref{eqn:defThmu} entering
the modular completion of the refined partition function 
$\widehat{h}_{p,\mu}$ reduces to the simple result
\be
\label{eqn:Th2ref}
\Theta_\mu( \{p_1,\mu_1,p_2,\mu_2\}; \tau,y) =\sum_{q_1+q_2=\mu+\frac12 p}
\frac{(-y)^{\gamma_{12}}}{2} 
\left( E_1\left(\frac{\sqrt{2\tau_2}(\gamma_{12}+\eta(pp_1p_2)) }{\sqrt{(pp_1p_2)}}\right)-\sign(\gamma_{12})\right)
e^{\I\pi\tau Q_2},
\ee 
where $\gamma_{12}=p_2 q_1-p_1 q_2$ and $(p p_1 p_2):=\kappa_{abc} p^a p_1^b p_2^c$. 
In the unrefined limit\footnote{Following  the prescription in 
\cite{Alexandrov:2019rth}, the unrefined limit
is obtained by setting $\eta=\frac{w}{2\I\tau_2}, \bar w=0$ and taking the limit  $w=\frac{\log y}{2\pi\I}\to 0$.},
 symmetrizing under the exchange of $q_1,q_2$ and using the identities $E_1(x)-\sign(x)=M_1(x)$, $\partial_x E_1(x)=e^{-\pi x^2}$, one arrives at 
\be
\label{eqn:Th2unref}
\Theta_\mu( \{p_1,\mu_1,p_2,\mu_2\}; \tau) =\sum_{q_1+q_2=\mu+\frac12 p}
\frac{(-1)^{\gamma_{12}}}{4} 
\left( \gamma_{12} \, M_1\left(\frac{\sqrt{2\tau_2} \gamma_{12} }{\sqrt{(pp_1p_2)}}\right)
+\frac{\sqrt{(p p_1 p_2)}}{\pi \sqrt{2\tau_2}} e^{-\frac{2\pi\tau_2 \gamma_{12}^2}{(p p_1 p_2)}} \right)
e^{\I\pi\tau Q_2}.
\ee
Analogous results for $n=3$ and $n=4$ will be discussed in Section \ref{sec:comparison}.

For later reference, we also recall the two-body contribution to the (refined) instanton generating potential (first found in \cite{Manschot:2010sxc}, rederived in \cite[Section  2.2]{Alexandrov:2016tnf}, and refined in \cite[(G.8)]{Alexandrov:2018lgp}), 
\be
\label{eqn:Th2G}
\begin{split}
\mathcal{G}_2 =& \sum_{p_1,p_2,q_1,q_2}
\frac{e^{2\pi\tau_2 \cI^2_{\gamma_1,\gamma_2}}}{8\pi^2 \sqrt{2\tau_2 (pt^2)}}
\left[
E_1\left(\sqrt{2\tau_2} \cI_{\gamma_1,\gamma_2} \right)
-\sign \gamma_{12}
\right]
e^{\I\pi\tau Q_2}\,
{h}_{p_1,q_1}\, {h}_{p_2,q_2}.
\end{split}
\ee
It is worth stressing that the argument of the error function in \eqref{eqn:Th2G} in general does 
{\it not} reduce to the argument of the error function in the modular completion \eqref{eqn:Th2ref} in the large volume attractor chamber, even for $\eta=0$. Indeed, $\cI_{\gamma_1,\gamma_2}$ 
 evaluates at that point to   
\be
\label{eqn:Istar}
\cI^*_{\gamma_1,\gamma_2} = \gamma_{12} \sqrt{\frac{(p^3)}{(p_1 p^2) (p_2 p^2)}} ,
\ee
such that the arguments of the two $E_1$ functions differ by a rescaling by 
 $\sqrt{ \frac{(p^3) (p_1 p_2 p)}{(p_1 p^2) (p_2 p^2)}}$. In the special case where the
D4-brane charges $p_1,p_2$ are collinear, this ratio evaluates to one and the two arguments are then identical.

\subsection{Generalized error functions}\label{sec:gen-erf}
Here we briefly review the definitions of the generalized error functions introduced in  \cite{Alexandrov:2016enp} and further developed in \cite{Nazaroglu:2016lmr}. 
For any invertible, $r\times r$ matrix $\cM$ with real entries, and vector $\vb{x}\in\IR^r$ we define 
the generalized error function $E_r$ and complementary error function $M_r$ by
\be
\label{eqn:defEgen}
  E_{r}(\mathcal{M}, \vb{x}) = \int_{\mathbb{R}^r} \dd[r]{\vb{z}} e^{-\pi(\vb{x}-\vb{z})^T(\vb{x}-\vb{z})} \prod_{i=1}^{r} \sgn (\mathcal{M}^{T} \vb{z})_i,
\ee
\be
\label{eqn:defMgen}
  M_r(\mathcal{M}; \vb{x}) = \qty(\frac{\I}{\pi})^{r} \qty|\det \mathcal{M}|^{-1} \int_{\mathbb{R}^r - \I \vb{x}} \dd[r]{\vb{z}} \frac{e^{-\pi \vb{z}^{T} \vb{z} - 2 \pi \I \vb{z}^T \vb{x}}}{\prod_{i = 1}^{r} \qty(\mathcal{M}^{-1} \vb{z})_i}.
\ee
We refer to the index $r$
 as the `depth'. The function $E_{r}(\mathcal{M}, \vb{x})$ is a $C^\infty$ function of $\vb{x}$, which asymptotes to 
$\prod_{i=1}^{r} \sgn (\mathcal{M}^{T} \vb{x})_i$ in the limit where $\vb{x}$ is scaled by an arbitrarily
large positive factor. Similarly $M_{r}(\mathcal{M}, \vb{x})$  is a $C^\infty$ function away from the hyperplanes $(\mathcal{M}^{-1} \vb{x})_i=0$, exponentially suppressed if $\vb{x}$ is scaled by an arbitrarily large positive factor. Moreover, both $E_r$ and $M_r$ are invariant (up to a sign $(-1)^r$) under $\vb{x}\mapsto - \vb{x}$, as well as 
under rescaling the columns of $\cM$ by arbitrary positive factors or permuting them, and under rotating $(\cM,\vb{x})\mapsto 
(\mathcal{O} \, \cM, \mathcal{O} \, \vb{x})$  by an arbitrary orthogonal transformation $\mathcal{O}\in O(r)$. For $r=1$, this allows
to set $\cM=1$, recovering the functions in \eqref{eqn:defE1},\eqref{eqn:defM1}. More generally, we can always reach a standard form such that $\cM$ is lower-triangular with unit diagonal. For displaying results with $r\leq 3$ we shall use the shorthand notations
\be\label{eqn:shorthand-gen-erf}
E_2(\alpha; x_1,x_2) = E_2\left( 
\begin{pmatrix} 1 & 0 \\ \alpha & 1 \end{pmatrix}, 
\begin{pmatrix} x_1 \\ x_2 \end{pmatrix} \right), \quad 
E_3(\alpha,\beta,\gamma; x_1,x_2,x_3) = E_3\left( 
\begin{pmatrix} 1 & 0 & 0 \\ \alpha & 1 & 0 \\ \beta & \gamma & 1 \end{pmatrix}, 
\begin{pmatrix} x_1 \\ x_2 \\x_3 \end{pmatrix} \right), \quad 
\ee
and similarly for $M_r$. 
The symmetry under permutations of the columns implies functional relations, for example \cite[Cor. 3.10]{Alexandrov:2016enp}
 \be\label{eqn:e2-basic-identities}
 E_2(\alpha; x_1,x_2) = E_2\left( \alpha; \frac{x_2-\alpha x_1}{\sqrt{1+\alpha^2}},
 \frac{x_1+\alpha x_2}{\sqrt{1+\alpha^2}}\right) = -E_2(-\alpha; -x_1, x_2),
 \ee
 and similar relations for $M_2$. In addition, every identity among products of sign functions 
 implies a corresponding identity among the generalized error functions $E_r(\mathcal{M}, \vb{x})$, where the matrices $\mathcal{M}$ are related by specific linear transformations  \cite[Eq.~(D.13)]{Alexandrov:2018lgp}. For example, the identity valid away from the loci where $x_1, x_2$ or $x_1+x_2$ vanish
\be
\sgn(x_1) \sgn(x_2) =   \left(\sgn(x_1) + \sgn(x_2)\right) \sgn(x_1 + x_2) -1 ,
\ee
implies the functional relation\footnote{See 
 \cite[Appendix  A.2]{Alexandrov:2024jnu} for more general functional relations.}
\be
 E_2(\mathcal{M}, \vb{x}) =   E_2(\mathcal{M}_1, \vb{x}) + E_2(\mathcal{M}_2, \vb{x})-1 ,
\ee
where the matrices $\mathcal{M}, \mathcal{M}_1, \mathcal{M}_2 \in \mathbb{R}^{2 \times 2}$
satisfy
\be
\label{eqn:E2M12}
  \mathcal{M} =
  \begin{pmatrix}
    \mathcal{M}_{11} & \mathcal{M}_{12} \\
    \mathcal{M}_{21} & \mathcal{M}_{22}
  \end{pmatrix},  \quad 
  \mathcal{M}_1 =
  \begin{pmatrix}
    \mathcal{M}_{11} & \mathcal{M}_{12} + \mathcal{M}_{11} \\
    \mathcal{M}_{21} & \mathcal{M}_{22} + \mathcal{M}_{21}
  \end{pmatrix}, \quad
  \mathcal{M}_2 =
  \begin{pmatrix}
    \mathcal{M}_{11} + \mathcal{M}_{12} & \mathcal{M}_{12} \\
    \mathcal{M}_{21} + \mathcal{M}_{22} & \mathcal{M}_{22}
  \end{pmatrix}.
\ee
In terms of the standard form defined in \eqref{eqn:shorthand-gen-erf}, the identity can be expressed as in a manifestly $\mathbb{Z}_3$-symmetric form, which will be relevant in the three-body case discussed in Section \ref{sec:results-four-centers} below,
\be
\label{eqn:E2trial}
  E_2(\alpha_1; x_1, y_1) + E_2(\alpha_2; x_2, y_2) + E_2(\alpha_3; x_3, y_3) = 1,
\ee
under the constraint
\be
  \alpha_1 \alpha_2 + \alpha_2 \alpha_3 + \alpha_3 \alpha_1 = 1,
\ee
provided the variables $(x_i,y_i)$ are related by 
\be
(x_{i+1},y_{i+1}) = \left( \frac{y_i+ \alpha_{i+1} x_i}{\sqrt{1+\alpha_{i+1}^2}}, 
 \frac{\alpha_{i+1} y_i-x_i}{\sqrt{1+\alpha_{i+1}^2}} \right),
\ee
where $\alpha_4:=\alpha_1$ and $(x_4,y_4)=\pm (x_1,y_1)$.

\medskip 

More generally, we need boosted versions of the generalized error functions, where the vector $\vb{x}$ lives in $\IR^d$ with fixed quadratic form $\cQ$ of signature $(r,d-r)$, and the arguments of the sign functions are given by a collection of $r$ vectors $(\vb*{v}_1,\dots,  \vb*{v}_r)$, corresponding to the columns of a $d\times r$ matrix $\cV$, 
spanning a positive definite subspace in $\IR^d$, \ie such that the Gram matrix $\cV^T \cQ \cV$ is positive definite. Choosing $\cB$ to be any $r\times d$ matrix whose rows define an orthonormal basis
for this subspace (hence $\cB \cQ \cB^T = 1$), we define the boosted generalized error functions as 
\be
\Phi^{E}_r(\cQ,\{ \vb*{v}_i\},   \vb{x} ) = E_r( \cB\cQ \cV, \cB \cQ \vb{x}), \quad 
\Phi^{M}_r(\cQ,\{ \vb*{v}_i\}, \vb{x} ) = M_r( \cB\cQ\cV, \cB \cQ \vb{x}).
\ee
Independence of the choice of basis follows from the $O(r)$ invariance of $E_r, M_r$. 
Similar to $E_r$ and $M_r$, 
$\Phi^E_r(\cQ,\cV, \vb{x} )$ is a $C^\infty$ function of $\vb{x}$ which asymptotes to\footnote{Here we denote $\sgn v:=\prod_{i =1}^r \sign v_i$ for a vector $v=(v_i)_{i=1}^r$.}
$\sgn( \cV^T \cQ \vb{x})$ as $\vb{x}$ is scaled to infinity, 
while $\Phi^M_r(\cQ,\cV, \vb{x} )$ is $C^\infty$ away from a set of $r$ hyperplanes 
and exponentially suppressed as $\vb{x}$ is scaled to infinity. In fact, $\Phi^{E}_r$ can be expressed in terms of $\Phi^{M}_r$ through the identity~\cite[Prop. 1]{Nazaroglu:2016lmr}, generalizing similar identities obtained in \cite{Alexandrov:2016enp} for $r=2,3$)
\be
\Phi^{E}_r(\cQ,\{ \vb*{v}_i\} , \vb{x} ) = \sum_{\cI\subset \{1,\dots r\} }
\Phi^{M}_{|\cI|} (\cQ, \{ \vb*{v}_i\}_{i\in \cI},   \vb{x} )  \prod_{j\notin \cI} 
\sign( \vb*{v}_{j\perp\cI} \cQ \vb{x}) ,
\ee
where $\vb*{v}_{j\perp\cI}$ is the projection of $\vb*{v}_j$ orthogonal to the vectors $\vb*{v}_i$ with $i\in \cI$, such that all discontinuities on the r.h.s. cancel, leaving a $C^{\infty}$ function of $\vb{x}$. For $n\leq 3$, this identity specializes to 
\begin{align}
\label{eqn:E1fromM}
E_1(x) &\,=\, M_1(x) + \sgn(x),  \\
\label{eqn:E2fromM}
E_2(\alpha, x_1, x_2) &\,=\, M_2\left(\alpha ;x_1,x_2\right)+M_1\left(\frac{\alpha  x_2+x_1}{\sqrt{\alpha ^2+1}}\right) \sgn\left(x_2-\alpha  x_1\right)+M_1\left(x_2\right) \sgn\left(x_1\right) \notag \\
&\quad + \sgn\left(x_2\right) \sgn\left(x_1+\alpha  x_2\right), \\
\begin{split}
\label{eqn:E3fromM}
  E_{3}\bigl(\alpha,\beta,\gamma;\,x_{1},x_{2},x_{3}\bigr) &\,=\, M_{3}\bigl(\alpha,\beta,\gamma;\,x_{1},x_{2},x_{3}\bigr)
+M_{2}\bigl(\gamma;\,x_{2},x_{3}\bigr)\,\sgn(x_{1})\\
&\quad+M_{2}\!\Bigl(\tfrac{\beta}{\sqrt{1+\alpha^{2}}};\,
   \tfrac{x_{1}+\alpha x_{2}}{\sqrt{1+\alpha^{2}}},\,x_{3}\Bigr)\,
   \sgn(-\alpha x_{1}+x_{2})\\
&\quad+M_{2}\!\Bigl(\tfrac{\alpha+\beta\gamma}
    {\sqrt{1+\beta^{2}-2\alpha\beta\gamma+(1+\alpha^{2})\gamma^{2}}};\;
   \tfrac{(1+\gamma^{2})x_{1}-(\beta-\alpha\gamma)(\gamma x_{2}-x_{3})}
        {\sqrt{(1+\gamma^{2})\bigl(1+\beta^{2}-2\alpha\beta\gamma+(1+\alpha^{2})\gamma^{2}\bigr)}},\;
   \tfrac{x_{2}+\gamma x_{3}}{\sqrt{1+\gamma^{2}}}
\Bigr)\, \\[-5pt]
   & \qquad \times \sgn\bigl({ (\alpha\gamma-\beta)x_{1}-\gamma x_{2}+x_{3}}\bigr)\\
&\quad+M_{1}(x_{3})\,\sgn(x_{2})\,\sgn(x_{1}+\alpha x_{2})\\
&\quad+M_{1}\!\Bigl(\tfrac{x_{2}+\gamma x_{3}}{\sqrt{1+\gamma^{2}}}\Bigr)\,
   \sgn\bigl({ (1+\gamma^{2})x_{1}-(\beta-\alpha\gamma)(\gamma x_{2}-x_{3}) }\bigr)\,
   \sgn(-\gamma x_{2}+x_{3})\\
&\quad+M_{1}\!\Bigl(\tfrac{x_{1}+\alpha x_{2}+\beta x_{3}}{\sqrt{1+\alpha^{2}+\beta^{2}}}\Bigr)\,
   \sgn\bigl({(1+\alpha^{2})x_{3}-\beta(x_{1}+\alpha x_{2})}\bigr)\, \\[-5pt]
   & \qquad \times\sgn\bigl({(1+\beta^{2}-\alpha\beta\gamma)\,x_{2}+(\gamma+\alpha^{2}\gamma-\alpha\beta)\,x_{3}-(\alpha+\beta\gamma)\,x_{1}}\bigr)\\
&\quad+\;\sgn(x_{3})\,\sgn(x_{1}+\alpha x_{2}+\beta x_{3})\,\sgn(x_{2}+\gamma x_{3})\,.
\end{split}
\end{align}

Functional relations originating from sign identities take a simple form in terms of boosted error functions, for example \eqref{eqn:E2trial} is equivalent to 
\be
  \Phi^E_2(\cQ,\{ \vb*{v}_1 + \vb*{v}_2, \vb*{v}_2 \}, \vb{x}) +
  \Phi^E_2(\cQ,\{ \vb*{v}_1, \vb*{v}_1 + \vb*{v}_2 \}, \vb{x}) =
  1 + \Phi^E_2(\cQ,\{ \vb*{v}_1, \vb*{v}_2 \}, \vb{x}).
\ee

\section{Witten index from localization}
\label{sec:witten-index}

In this section, we evaluate the Witten index of the supersymmetric quantum mechanics of
$n$ dyonic black holes using localization. We start by recalling the Lagrangian, and the structure of the supersymmetric ground states, both at the semiclassical and quantum level. Readers familiar with this may skip to the localization computation in Section \ref{sec:computing-witten-index}. 

\subsection{Supersymmetric quantum mechanics of \texorpdfstring{$n$}{n} dyons}

Following \cite{Denef:2002ru} (see also \cite{Lee:2011ph,Kim:2011sc}), the quantum mechanics of $n$ BPS dyons of charge 
$\{\gamma_i\}_{i=1}^n$
in  3+1-dimensional theories 
with $\mathcal{N}=2$ supersymmetry is given by the Lagrangian 
\be
  L = \sum_{i=1}^n \frac{m_i}{2} \left( \dot{\va{x}}_i^2 + D_i^2 + 
  2\I\bar{\lambda}_i\dot{\lambda}_i \right) + \sum_{i=1}^{n} (-U_i D_i + \va{A}_i \cdot \dot{\va{x}}_i) + \sum_{i, j = 1}^{n}  \va{\nabla}_i U_j \cdot \bar{\lambda}_i \va{\sigma}\lambda_j,
\ee 
where $\va{x}_i$ denotes the location of the $i$-th dyon of mass $m_i$ in $\IR^3$, $\lambda_i$ 
are the associated fermionic zero-modes, $D_i$ are auxiliary fields and $\va{\sigma}$
are the standard Pauli matrices. The functions $U_i(\va{x})$ 
and 3-vectors $\va{A}_i(\va{x})$ are determined from the DSZ pairings
$\gamma_{i j}=\langle \gamma_i, \gamma_j \rangle$ and FI parameters 
$c_i$ (subject to the constraint $\sum_i c_i=0$) via 
\be
  U_i = -\frac{1}{2}\qty(\sum_{j \neq i} \frac{\gamma_{i j}}{|\va{x}_i - \va{x}_j|} - c_i), \quad 
   \va{\nabla}_i U_j = \va{\nabla}_j U_i 
   = \frac{1}{2} (\va{\nabla}_i \times \va{A}_j + \va{\nabla}_j \times \va{A}_i).
\ee
 This Lagrangian admits $4$ real supersymmetries (corresponding to the unbroken supercharges in four dimensions), and  follows from the supersymmetric quiver quantum mechanics with $n$ Abelian vector multiplets and $|\gamma_{i j}|$ bifundamental chiral multiplets for each pair $(i,j)$, after integrating out the chiral multiplets \cite{Denef:2002ru}. 
Integrating out the auxiliary fields $D_i$ leads to a scalar potential 
\be
V=\sum_{i=1}^n \frac{U_i^2}{2 m_i}.
\ee
We denote by $\cM_n(\{\gamma_i, c_i\})$ the space of classical 
supersymmetric ground states up to overall translations, 
\be
\label{eqn:defMDenef}
\cM_n(\{\gamma_i, c_i\}) = 
\left\{  (\va{x}_i) \in \IR^{3n},\ 
\forall i\  \sum_{j \neq i}\frac{\gamma_{i j}}{|\va{x}_i - \va{x}_j|} = c_i 
\right\} 
 \bigg/ \IR^3.
\ee
As observed in \cite{Denef:2002ru},  
this is exactly the space of multi-centered BPS black hole solutions in $\cN=2$, four-dimensional supergravity if one identifies 
\be\label{eqn:FI-BPS}
c_i = 2 \Im(e^{-i \phi} Z(\gamma_i)), \quad m_i = |Z(\gamma_i)|, \quad \phi = \arg(Z(\gamma)),
\ee
with all central charges evaluated at spatial infinity. With these identifications, the asymptotic value of the potential $V_{\infty}=\sum_{i=1}^n \frac{c_i^2}{8m}$ coincides with the binding energy $\sum_{i=1}^n |Z_i| - |Z|$, in the limit where all central charges $Z_i$ are nearly aligned. This is indeed the case for D4-branes in the large volume limit, where $Z_i \simeq -\frac12 (p_i t^2)\in \IR^-$.

\subsection{Phase space of multicentered BPS black holes and Dirac index}
\label{sec:phase-space}

Since the $n$ constraints $U_i=0$ sum up to zero, the space 
of multicentered BPS black holes defined in \eqref{eqn:defMDenef} has even real dimension $2n-2$.
As noted in \cite{deBoer:2008zn},  it carries a natural symplectic form 
\be
\label{eqn:defom}
 \omega = \frac{1}{2} \sum_{i, j = 1}^{n} \epsilon_{a b c} \ \nabla^{a}_j U_i \, \de{\mathrm{x}}_i^b \wedge \de{\mathrm{x}}_j^c
 = \frac{1}{4} \sum_{i < j} \epsilon_{a b c} \frac{\gamma_{i j}}{r_{i j}^3} \, \mathrm{x}_{i j}^a \, \de{\mathrm{x}}_{i j}^b \wedge \de{\mathrm{x}}_{i j}^c,
 \ee
 and a Hamiltonian action of $SO(3)$ with moment map
 \be
 \label{eqn:defJ}
 \va{J} = \frac12 \sum_{i < j}  \gamma_{i j}\, \frac{\va{x}_{i j}}{r_{i j}} ,
 \ee
which coincides with the total angular momentum carried by the multi-centered black hole solution. 

When the phase space $\cM_n(\{\gamma_i, c_i\})$ is compact\footnote{This is the case when the corresponding quiver has no closed loops, or more generally when scaling solutions are absent, which depends
on certain linear inequalities on the  $\gamma_{i j}$'s  \cite{Bena:2012hf,Descombes:2021egc}.}, its equivariant volume
\be
\label{eqn:defgclass}
\Vol(\{\gamma, c\}; \, y) = \frac{ (-1)^{\sum_{i < j} \gamma_{ij} -n+1}}
{(2\pi)^{n-1} (n-1)!} \int_{\cM_n} y^{2\mathrm{J}_3} \omega^{n-1},
\ee
can be computed using the Duistermaat-Heckmann formula \cite{Manschot:2010qz}: the fixed points with respect to rotations along the $z$ axis (or equivalently, the critical points of 
$\mathrm{J}_3$) are collinear configurations $\va{x}_i=(0,0,z_i)$
subject to the one dimensional reduction of the equations $U_i=0$. They are in general isolated,
and characterized by the order of the centers along the $z$ axis. We denote by $F_{C,n}( \{ \gamma_i, c_i\})$ the number of solutions with the standard ordering $z_1<z_2<\dots < z_n$, weighted by 
the sign of the determinant of the Hessian matrix of the `superpotential' $W=-\sum_{i<j} \gamma_{ij} \log|z_j-z_i| - \sum_i c_i z_i$. The equivariant
volume is then given by a sum over all permutations, 
\be
\Vol(\{\gamma, c\}; \, y) = \frac{ (-1)^{\sum_{i<j} \gamma_{ij} -n+1}}
{(2\log y)^{n-1}}  \sum_{\sigma\in S_n}
 F_{C, n} (\{ \gamma_{\sigma(i)}, c_{\sigma(i)}\}) \, y^{\sum_{i < j} \gamma_{\sigma(i) \sigma(j)}}.
\ee
Quantum mechanically, the index counting normalizable BPS states is expected to coincide\footnote{Possibly up to an overall sign, depending whether the harmonic spinors are supported on positive or negative chirality \cite[Section 2.5]{Manschot:2011xc}.}
 with the equivariant index of the Dirac operator twisted by a line bundle whose first Chern class is equal to the symplectic form $\omega$ \cite{deBoer:2008zn}. The latter can be computed by replacing
 the volume form  in \eqref{eqn:defgclass} by a suitable equivariant cohomology class
 \be
 \label{eqn:Aroofrep}
  \frac{\omega^{n-1}}{(2\pi)^{n-1}} 
   \rightarrow  \left[ 
    e^{\frac{\omega}{2\pi}} 
\widehat{A}(\cM_n,y) 
   \right]_{n-1},
 \ee
 where $ \widehat{A}(\cM_n,y)$ is the equivariant $A$-roof genus of $\cM_n$, 
 and $[A]_{n-1}$ denotes the degree $2n-2$ part of the multiform $A$ (see \cite[Section 2.5]{Manschot:2011xc} for more details). This integral can in turn
 be computed  using the Atiyah-Bott Lefschetz fixed point theorem \cite{Manschot:2011xc}, and (because the fixed points are isolated) it coincides with the equivariant volume up to a simple rescaling, arising from the $A$-roof genus at the fixed points, 
\be
\label{eqn:gref}
\begin{split}
\Ind_C(\{\gamma, c\}; \, y) =& \left( \frac{2\log y}{y-1/y} \right)^{n-1}\!\!\!\!
\Vol(\{\gamma, c\}; \, y) \\
& = \frac{ (-1)^{\sum_{i<j} \gamma_{ij} -n+1}}
{(y-1/y)^{n-1}} \sum_{\sigma\in S_n}
  F_{C, n} (\{ \gamma_{\sigma(i)}, c_{\sigma(i)}\}) \, y^{\sum_{i < j} \gamma_{\sigma(i) \sigma(j)}}.
\end{split}
\ee

Now, if the phase space $\cM_n$ is compact, the r.h.s. of \eqref{eqn:gref} is guaranteed to be a symmetric Laurent polynomial in $y$, and correctly computes the character of the $SO(3)$ action on the space of 
harmonic spinors. In the presence of scaling solutions, however, $\cM_n$ is no longer compact, and 
the r.h.s. of  \eqref{eqn:gref} is in general a rational function of $y$, which cannot be understood as a $SO(3)$ character. A possible remedy, advocated in \cite{Manschot:2011xc}, is to include suitable 
boundary contributions such that the final result is a symmetric Laurent polynomial with integer coefficients. This procedure was designed to extract  single-centered black hole indices, whose mathematical definition has remained elusive. 

Instead, for the purposes of extracting attractor indices (of which MSW indices are a special case), the correct procedure is to use the flow tree formula, which relies on a partition of the phase space of multi-centered black holes into components labeled by attractor flow trees. 
As shown in \cite{Alexandrov:2018iao}, this prescription leads to a similar formula as in \eqref{eqn:gref},
\be
\label{eqn:gtree}
\Ind_\mathrm{tree}(\{\gamma, c\}; \, y) = \frac{ (-1)^{\sum_{i<j} \gamma_{ij} -n+1}}
{(y-1/y)^{n-1}} \sum_{\sigma\in S_n}  F_{\mathrm{tree}, n} (\{ \gamma_{\sigma(i)}, c_{\sigma(i)}\}) \, y^{\sum_{i < j} \gamma_{\sigma(i) \sigma(j)}},
\ee
where the coefficients $F_{\mathrm{tree}, n} (\{ \gamma_i, c_i\})$ (sometimes called `partial tree indices') now count planar attractor flow trees. Those can be computed recursively as follows~\cite{Alexandrov:2018iao}:

%

\be\label{eqn:flow-tree-recursion}
F_{\text{tree},n}(\{\gamma, c\}) = F_n^{(0)}(c_1, \, \dots, \, c_n) 
- \sum_{\substack{n_1 + \cdots + n_m = n \\ n_k \geq 1,\; m < n}} 
F_{\text{tree},m}(\{\gamma', c'\}) \prod_{k=1}^{m} F_{n_k}^{(0)}(c^{\star}_1, \, \dots, \, c^{\star}_{n_k}),
\ee
where $F^{(0)}$ is given by
\be
  F_n^{(0)}(c_1, \, \dots, \, c_n) = \frac{1}{2^{n-1}} \prod_{i=1}^{n-1} \sgn(S_i), \qquad S_i = c_1 + c_2 + \ldots + c_i.
\ee
The sum in \eqref{eqn:flow-tree-recursion} runs over all \textit{compositions} (ordered partitions) of $n$ with length $m < n$%
\footnote{One can compute the number of compositions, $p_{\rm C}(n)$ as follows: write $n$ as a row of $n$ ones and make an independent binary choice in each of the $n-1$ interstices: either insert a separator (starting a new part) or leave the gap open.  This produces $2^{n-1}$ compositions; the single pattern that inserts separators everywhere gives the length-$n$ composition $(1,\dots ,1)$, which is excluded.  Hence $p_{\rm C}(n)=2^{n-1}-1$.}.
For a particular composition $(n_1, \ldots, n_m)$, we define the list of $m$ charges
and FI parameters
\be
  \gamma'_{k} = \sum_{\ell = 1}^{n_k} \gamma_{s_{k-1} + \ell}, \qquad c'_{k} = \sum_{\ell = 1}^{n_k} c_{s_{k-1} + \ell}, \qquad k = 1, \, \dots, \, m.
\ee
where $s_k = n_1 + \dots + n_k$ and $s_0 = 0$.  
For a particular charge $\gamma'_{k} = \sum_{\ell = 1}^{n_k} \gamma_{s_{k-1} + \ell}$, we further define the $n_k$ numbers $c^{\star}_\ell$ as the attractor value of the FI parameters for this subset, \ie $c^{\star}_\ell = \langle \gamma'_k, \gamma_{s_{k-1} + \ell} \rangle$. The initial value for the recursion is set by $F_{\mathrm{tree}, 1}(\{\gamma_1,c_1\}) = 1$.

As an illustration of this recursion relation, consider the partial tree index for two charges $(\gamma_1, \gamma_2)$ for FI parameters $(c_1, c_2)$ such that $c_1 + c_2 = 0$. There is only one composition $(2)$ and for this composition $\gamma' = \gamma_1 + \gamma_2$ and $c^{\star}_1 = -\gamma_{1 2}$. Thus,
\be
  F_{\mathrm{tree}, 2} = \frac{1}{2} \sgn(c_1) - \frac{1}{2} \sgn(c_1^{\star}) = \frac{1}{2} \qty\Big[ \sgn(c_1) + \sgn(\gamma_{1 2}) ].
\ee
For three centers, there are $3$ compositions $(1, 2)$, $(2, 1)$, and $(3)$ and we get
\be
  \begin{aligned}
    F_{\mathrm{tree}, 3} &= F^{(0)}_{3}(c_1, c_2, c_3) - F_{\mathrm{tree}, 2}\qty(\gamma_1, \gamma_{2 + 3}) F^{(0)}_{2}(-\gamma_{2 3}, \gamma_{2 3}) - F_{\mathrm{tree}, 2}\qty(\gamma_{1 + 2}, \gamma_{3}) F^{(0)}_{2}(-\gamma_{1 2}, \gamma_{1 2}) \\
    &\quad - F^{(0)}_{3} (\gamma_{2+3, 1}, \gamma_{1+3, 2}, \gamma_{1+2, 3}),
  \end{aligned}
\ee
where $\gamma_{2+3, 1}:=\gamma_{21}+\gamma_{31}$, and similarly for $\gamma_{1+3, 2}$.
The full result for up to four centers was given in \cite[Section C.1]{Alexandrov:2018iao}, and will be recalled when we need it in Section \ref{sec:results-four-centers} below.

\subsection{Computing the Witten index by localization }
\label{sec:computing-witten-index}

\medskip

In the remainder of this section, our goal is to evaluate the refined Witten index of this supersymmetric quantum mechanics, which includes the contribution from the whole spectrum, rather than just normalizable bound states. The key observation is that, for non-linear sigma models with potential\footnote{The occurrence of terms at first-order in time derivatives does not affect this conclusion.}, the  functional integral computing the Witten index reduces to an integral over constant modes 
\cite{Girardello:1983pw,Imbimbo:1983dg}. Thus, we are left to evaluate the following integral,\footnote{The factor of $\I$ in the exponent arises due to Wick rotation $D\to -\I D$. The normalization
factor $(4\pi^2\beta)^n$ will be justified a posteriori by matching the contribution of normalizable bound states.}
\be
\label{eqn:genIn}
\cI_{n}(\{\gamma, c\},\beta)
=  \int \prod_{i=1}^{n} 
\frac{  \dd[3]{\va{x}}_i\, 
  \de \bar \lambda_i\, 
  \de \lambda_i\,
  \de D_i}{4\pi^2\beta}\, 
 \exp\left(-\beta \qty( \sum_{i=1}^{n} \qty(\I U_i D_i + \frac{m_i}{2} D_i^2) + \sum_{i, j=1}^n  \va{\nabla}_j{U_i} \bar{\lambda}_i \va{\sigma}\lambda_j ) \right).
\ee
Formally this integral is infinite due to the flat directions corresponding to overall translations of the position multiplet: $(\va{x}_i, \lambda_i, \bar{\lambda}_i, D_i) \to (\va{x}_i + \va{d}, \lambda_i + \chi, \bar{\lambda}_i + \bar{\chi}, D_i + t)$. To remove these \textit{center of mass} degrees of freedom, we can do the following change of variables
\be
  \{\va{x}_i\}_{i=1}^{n} \ \longrightarrow \ \{\tilde{\va{x}}_i = \va{x}_i - \va{x}_n \}_{i = 1}^{n-1}, \quad \tilde{\va{x}}_0 = \frac{\sum_i m_i \va{x}_i}{\sum_i m_i},
\ee
and similarly for $\tilde{\lambda}_i$, $\bar{\tilde{\lambda}}_i$, and $\tilde{D}_i$. We immediately see the following relations
\be
  \sum_{i=1}^n U_i D_i = \sum_{i=1}^{n-1} U_i \tilde{D}_i, \quad \sum_{i, j = 1}^{n}  
  \va{\nabla}_j{U_i(\va{x})} \, \bar{\lambda}_i \va{\sigma} \lambda_j = \sum_{i, j = 1}^{n-1} 
   \va{\nabla}_j{U_i(\tilde{\va{x}})} \, \bar{\tilde{\lambda}}_i \va{\sigma} \tilde{\lambda}_j.
\ee
using $\sum_i U_i = 0$ and the fact that $U_i(\va{x})$ is translation invariant. The quadratic term for the auxiliary fields becomes
\be
\label{eqn:defM}
  \sum_{i=1}^{n} \frac{m_i}{2} D_i^2 = \frac{m_\mathrm{tot}}{2} \tilde{D}_0^2 + \frac{1}{2} \sum_{i, j = 1}^{n-1} \sfM_{i j} \tilde{D}_i \tilde{D}_j, \quad \text{where } \quad \sfM_{i j} = m_i \delta_{i j} - \frac{m_i m_j}{m_\mathrm{tot}}. 
\ee
We shall refer to the $(n-1) \times (n-1)$ matrix $\sfM$ as the reduced mass matrix. For example, for $n = 2$ bodies, $M_{1 1} = \frac{m_1 m_2}{m_1 + m_2}$ is the usual reduced mass. Thus, the integral along the flat direction in \eqref{eqn:genIn} can be isolated as $\frac{1}{(2\pi)^2 \beta}\int \dd[3]{\tilde{\va{x}}_0 \dd{\tilde{\lambda}_0} \dd{\bar{\tilde{\lambda}}_0} \dd{\tilde{D}_0}} e^{-\beta \frac{m_\mathrm{tot}}{2} \tilde{D}_0^2}$. Having removed the center of mass direction, let us rewrite \eqref{eqn:genIn}. For brevity, we will not put tilde's to denote the relative coordinates but instead use the same notation as the original coordinates,
\be
\label{eqn:genInRed}
\cI_{n}(\{\gamma, c\},\beta)
=  \int \prod_{i=1}^{n-1} 
\frac{  
 \dd[3]\va{x}_i\, 
  \de \bar \lambda_i\, 
  \de \lambda_i\,
  \de D_i}{4 \pi^2 \beta}\, 
 \exp\left(-\beta \qty( \sum_{i=1}^{n-1} \I U_i D_i + \sum_{i, j=1}^{n-1} \qty(\frac{\sfM_{i j}}{2} D_i D_j +  \va{\nabla}_j{U_i} \bar{\lambda}_i \va{\sigma}\lambda_j) ) \right).
\ee
Similarly, the refined index is expected to be computed by the integral
\be
\label{eqn:genIny}
\begin{split}
\cI_{n}(\{\gamma, c\},\beta,y)
= &\left( \frac{2\log y}{y-1/y}\right)^{n-1}
 \int \prod_{i=1}^{n-1} 
\frac{  \dd[3] \va{x}_i \, 
  \de \bar \lambda_i\, 
  \de \lambda_i\,
  \de D_i}{4\pi^2\beta} \\ &
  y^{2J_3}
 \exp\left(-\beta \qty( \sum_{i=1}^{n-1} \I U_i D_i + \sum_{i, j=1}^{n-1}  \qty(\frac{1}{2}\sfM_{i j} D_i D_j + \va{\nabla}_j{U_i} \bar{\lambda}_i \va{\sigma} \lambda_j)) \right),
 \end{split}
\ee
where $J_3$ is the $z$-component of the angular momentum defined in \eqref{eqn:defJ}, and the power of 
$2\log y/(y-1/y)$ is motivated by the relation \eqref{eqn:gref} between equivariant volume and equivariant Dirac index.

\subsubsection{Two-body case, revisited}\label{sec:two-body-case}
As a warm-up, we consider the two-body case, which was already studied at length in \cite{Pioline:2015wza}. For brevity, we denote $\va{x}=\va{x}_{12}, \kappa=\gamma_{12}, c_1=-c_2=c$ 
and define the reduced mass 
\be 
\label{eqn:redmass}
m = m_{12} := \frac{m_1 m_2}{m_1 + m_2}.
\ee
Focusing first on the unrefined index, 
we have to evaluate
\be
\label{eqn:I2unref}
\cI_2= \frac{1}{4\pi^2\beta} \int \de^3 \va{x} \ \de D \ \de^2 \lambda\, \de^2 \bar{\lambda} \ e^{-\beta \left( \I U D + \frac{m}{2}D^2 + \va{\nabla} U \cdot \bar{\lambda} \va{\sigma} \lambda \right)},
\ee
where $U = -\frac{1}{2} \qty(\kappa/|\va{x}| - c)$. The fermion contribution readily evaluates to
\be
 \int \de^2 \lambda \ \de^2 \bar{\lambda} \ e^{-\beta \va{\nabla} U \cdot \bar{\lambda} \va{\sigma} \lambda} = -\beta^2 (\va{\nabla} U)^2 = -\frac{\beta^2 \kappa^2}{4 |\va{x}|^4}.
\ee
Going to spherical coordinates, the integral becomes
\be
\label{eqn:I2sph}
  \begin{aligned}    
    \cI_2 &= -\frac{\beta \kappa^2}{4 \pi} \int_0^\infty  \frac{r^2\de r}{r^4} \int_{-\infty}^\infty \de D \ e^{-\beta \left( -\frac{\I}{2} D \left( \frac{\kappa}{r} - c\right) + \frac{m}{2}D^2 \right)}\\
    &= -\frac{\beta \kappa^2}{4 \pi} \left( \frac{\beta m}{2 \pi} \right)^{-\frac{1}{2}} \int_0^\infty \de \rho \ e^{-\frac{\beta}{8 m} (\kappa \rho - c)^2},
  \end{aligned}
\ee
where we defined $\rho=1/r$ and integrated over $D$ in the second line. We can compute 
the integral by extending its range 
to the full real axis by inserting the step function $\frac{\sign \rho + 1}{2}$:
\bea
  \int_0^\infty \de \rho \ e^{-\frac{\beta}{8 m} (\kappa \rho - c)^2} &=& 
\frac{1}{2} \int_{-\infty}^\infty \de \rho \ e^{-\frac{\beta}{8 m} (\kappa \rho - c)^2}
+ \frac{1}{2}\int_{-\infty}^\infty \de \rho \ \sign (\rho) \ e^{-\frac{\beta}{8 m} (\kappa \rho - c)^2} \nn\\
&=& \frac{1}{2 }\left(\frac{\beta \kappa^2}{8 \pi m}\right)^{-\frac{1}{2}} + 
\frac{1}{2}\left(\frac{\beta \kappa^2}{8 \pi m}\right)^{-\frac{1}{2}} 
\int_{-\infty}^\infty \de u \ \sign (u) \ e^{-\pi \left(u - \sqrt{\frac{\beta}{8\pi m}} \sgn(\kappa) c \right)^2} \nn \\
&=& \frac{1}{2}\left(\frac{\beta \kappa^2}{8 \pi m}\right)^{-\frac{1}{2}}   \left[ 1 + \sign(\kappa)\, 
E_1 \left( c \, \sqrt{\frac{\beta}{8 \pi m}}\right) \right],
\eea
where we changed variable to $u = \sqrt{\frac{\beta}{8 \pi m}} |\kappa| \rho$
in the second line and used the definition \eqref{eqn:defE1} of the error function.  Substituting in 
\eqref{eqn:I2sph}, we arrive at 
\be
\label{eqn:I2final}
\begin{split}
\cI_2=& -\frac{\kappa}{2} \left[ \sign(\kappa) + E_1\left(  c \,  \sqrt{\frac{\beta}{8\pi m}} \right) \right] \\
=& -\frac{\kappa}{2} \left[ \sign(c) + \sign(\kappa) + M_1\left( c \,  \sqrt{\frac{\beta}{8\pi m}} \right) \right].
\end{split}
\ee
\begin{figure}[htbp]
  \centering
  \begin{subfigure}{0.31\textwidth}
      \centering
      \includegraphics[width=136pt]{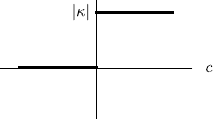}
      \caption{Contribution from supersymmetric bound states, proportional to the step function $\sign(c)$ and equal to $|\kappa|$ when $c$ and $\kappa$ have the same sign.}
      \label{fig:two-body-index-Bound}
  \end{subfigure}
  \hfill
  \begin{subfigure}{0.31\textwidth}
      \centering
      \includegraphics[width=136pt]{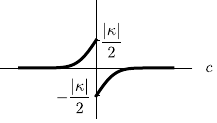}
      \caption{Contribution from the continuum of scattering states, given by the complementary error function $M_{1}(c)$ and interpolating discontinuously between $\pm |\kappa|/2$.}
      \label{fig:two-body-index-Scatt}
  \end{subfigure}
  \hfill
  \raisebox{13pt}{
  \begin{subfigure}{0.31\textwidth}
      \centering
      \includegraphics[width=136pt]{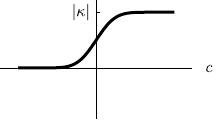}
      \caption{Total two-body Witten index $\mathcal{I}_{2}=\mathcal{I}_{2}^{(\mathrm{bound})}+\mathcal{I}_{2}^{(\mathrm{scatt})}$, a smooth error function completion across the wall of marginal stability.}
      \label{fig:two-body-index-total}
  \end{subfigure}
  }
  \caption{Decomposition of the two-body Witten index into contributions from bound states (a) and scattering states (b). Their sum (c) produces a smooth interpolation described by the error function $E_{1}(c)$, consistent with the modular completion of the BPS index.}
  \label{fig:two-body-index}
\end{figure}
The expression in the first line shows that the Witten index is a smooth function of the stability parameter $c$, as emphasized in \cite{Pioline:2015wza}. The second line expresses it, as the sum of the contributions of the $|\kappa|$ supersymmetric bound states, 
which exist whenever $c$ and $\kappa$ have the same sign, and the contribution of the continuum of scattering states, proportional to the complementary error function $M_1$. Indeed, the latter can be rewritten as 
\be
 M_1\left( c \,  \sqrt{\frac{\beta}{8\pi m}} \right) = 
 -\frac{2c}{\pi} \int_{2|c|}^{\infty} \frac{\de k}{k \sqrt{4k^2-c^2}} e^{-\frac{\beta k^2}{2m}},
\ee
consistent with the spectral asymmetry computed in \cite[(3.36)]{Pioline:2015wza} (where $k$ is the norm of the momentum in $\IR^3$). This decomposition is illustrated in \cref{fig:two-body-index}. In terms of the 
combination $\cI_{\gamma_1,\gamma_2}$ defined in \eqref{eqn:ImZ1Z2}
and of the modular parameter $\tau_2=\beta/(2\pi)$, the result \eqref{eqn:I2final} may be rewritten as
\be
\cI_2=-\frac{\gamma_{12}}{2} 
\left[  \sign(\gamma_{12}) - E_1\left( \sqrt{2\tau_2} \cI_{\gamma_1,\gamma_2}\right) \right].
\ee
Comparing with two-body contribution to the instanton generating potential \eqref{eqn:Th2G}, we find perfect agreement up to overall normalization factors.
However, if one specializes to the large volume attractor chamber $c =2\lambda \gamma_{12}$, leading to the $\lambda$-independent ratio $\cI^*_{\gamma_1,\gamma_2}$ in \eqref{eqn:Istar},  we find agreement with the modular completion \eqref{eqn:Th2unref}
only up to a rescaling of the argument of the complementary error function by a factor $\sqrt{\frac{(p p_1 p_2)(p^3)}{(p_1 p^2) (p_2 p^2)}}$.  At present we do not understand the physical origin of this rescaling. Nevertheless, we proceed since it disappears when the D4-brane charges are collinear, which covers the case of one-parameter compact CY threefolds as well as non-compact CY threefolds of the form $X=K_S$ for $S$ a complex surface.

\medskip

A second puzzle is that 
the localization computation does not reproduce the Gaussian term in \eqref{eqn:Th2unref}. This term is necessary for modular invariance, and arises by differentiation from 
the refined completion \eqref{eqn:Th2unref}, due to the $\eta$-dependence in the argument.
One may attempt instead to compute the refined modular completion \eqref{eqn:Th2ref} directly by inserting the factor
$y^{2J_3}$ in the integral, as in \eqref{eqn:genIny}. The computation proceeds as before, except that the integral over the
angular coordinates now includes a factor $y^{\kappa \cos\theta}$, which replaces the $4\pi$ 
solid angle by
\be
4\pi \mapsto \int_{0}^{2\pi} \de\phi\,  \int_{-\pi}^{\pi}  \sin\theta\, \de \theta \, y^{\kappa \cos\theta}
= \frac{2\pi}{\kappa\log y} \left( y^{\kappa}-y^{-\kappa} \right).
\ee
We thus obtain
\be\label{eqn:two-body-result-refined}
\cI_2(y) = -\frac{1}{2}
\left( \frac{ y^{\kappa}-y^{-\kappa} }{y-1/y}  \right)
\left[ \sign(\kappa) + E_1\left(  c \,  \sqrt{\frac{\beta}{8\pi m}} \right) \right],
\ee
in precise agreement with the refined modular completion \eqref{eqn:Th2ref} under
the same identifications as above, provided $\eta=0$, \ie $y$ is a pure phase. The microscopic
effect of the parameter $\eta$, which is  connected to the Gaussian term in  \eqref{eqn:Th2unref},
unfortunately remains mysterious at this point.  

\medskip

\begin{figure}[htbp]
  \centering
  \begin{subfigure}[t]{0.45\textwidth}
      \centering
      \includegraphics[width=\linewidth]{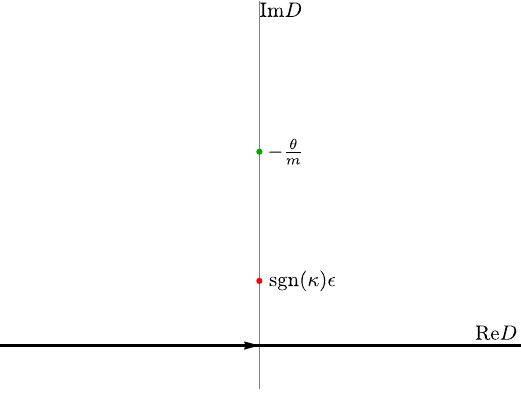}
      \caption{The original contour integral in the auxiliary field $D$ along with the pole at $D = i \sgn(\kappa)\epsilon$ (in red) and the saddle point of the integrand, $D = -i \frac{\theta}{m}$ (in green). This figure is for the case when $\kappa \theta < 0$.}
      \label{fig:contour-integral-pre}
  \end{subfigure}
  \hfill
  \begin{subfigure}[t]{0.45\textwidth}
      \centering
      \includegraphics[width=\linewidth]{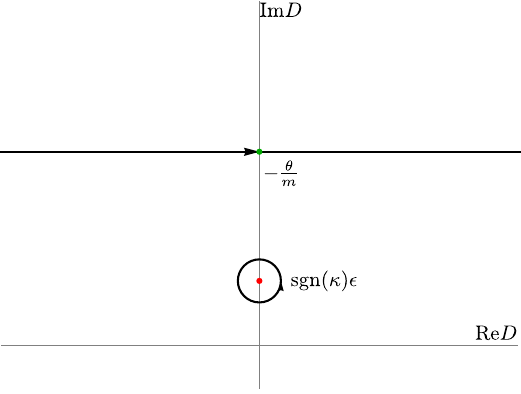}
      \caption{After shifting the contour to pass through the saddle, $D = -i\theta/m$, we pick up the contribution from the pole at $D = i \sgn(\kappa) \epsilon$ coming from the bound state in this case.}
      \label{fig:contour-integral-post}
  \end{subfigure}
  \caption{Contour integral in the $D$ plane}
  \label{fig:contour-integral}
\end{figure}

Finally, it is instructive to compute the same integral without eliminating the auxiliary fields. 
For this, we must be more specific about the contour of integration. We claim that the correct choice
is to shift $D\mapsto D-\I\epsilon\,\sign \kappa$, where $\epsilon>0$ is an infinitesimal positive parameter. This ensures that the integral over $\rho=1/r$ in the first line of \eqref{eqn:I2sph} is convergent
and becomes
\be
\cI_2 =  \frac{\I \kappa}{2\pi} 
\int_{-\infty}^\infty \frac{\de D}{D - \I \epsilon \sgn\kappa} \ e^{-\frac{\beta}{2} \left(\I c D  + 
m D^2  \right)}.
\ee
Upon shifting the contour of $D$ integration to run across the saddle point $D_* = -\frac{\I c}{2m}$, we recognize the integral representation \eqref{eqn:defM1} of the complementary error function,
\be
\label{eqn:I2saddle}
\cI^{\rm saddle}_2 = \frac{\kappa}{2}\, 
  M_1\left( c\, \sqrt{\frac{\beta}{8\pi m}}  \right).
\ee
However, we have to take into account the pole at $D = \I \epsilon \sgn\kappa $. Clearly, in shifting the contour we encounter the pole only if $\kappa$ has the same sign as $c$ (see Fig. \ref{fig:contour-integral}). If this is the case, we pick up an additional pole contribution 
\be
\label{eqn:I2pole}
    \cI^\mathrm{pole}_2 = -\kappa \, \sgn(\kappa) \Res_{D = \I \epsilon \sgn\kappa } \left( 
    \frac{e^{-\frac{\beta}{2} \left(\I c D + m D^2  \right)}}
    {D - \I \sgn(\kappa) \epsilon} \right)
    = -|\kappa|,
\ee
where the factor of $\sgn(\kappa)$ before the residue comes from the fact that the contour closes counterclockwise if $\kappa > 0$ and clockwise if $\kappa < 0$. It is easy to see that 
the sum of \eqref{eqn:I2saddle} and \eqref{eqn:I2pole} reproduces \eqref{eqn:I2final}.


\subsubsection{General case}\label{sec:general_case}

We now turn to the general $n$-body Witten index defined in \eqref{eqn:genInRed}. The fermion integral gives the determinant
\be
\int \prod_{i=1}^{n-1} 
  \de \bar \lambda_i\, 
  \de \lambda_i\,
 \exp\left(-\beta  \sum_{i, j=1}^{n-1}  \va{\nabla}_j U_i\, \bar{\lambda}_i \va{\sigma}\lambda_j \right) 
 =  (\beta^2)^{n-1} \det(\va{\nabla}_j U_i \otimes \va{\sigma}),
\ee
The key observation is that the space of relative positions $\IR^{3n-3}$ is foliated
by the phase spaces $\cM_n(\{\gamma_i,u_i\})$ introduced in \eqref{eqn:defMDenef}, with arbitrary 
FI parameters $u_i\in \IR^{n}$ (with $\sum_i u_i=0$), such that the flat measure on 
$\IR^{3n-3}$ times the fermion determinant equals the flat measure on $\IR^{n-1}$ times the symplectic
volume form $\omega^{n-1}$ on each leaf:
\be\label{eqn:master-formula}
\boxed{  \prod_{i=1}^{n-1} \de^3{\va{x}_{i}} \det(\va{\nabla}_i U_j \otimes \va{\sigma}) = 
   \frac{(-1)^{n-1}}{2^{n-1}(n-1)!}
   \left(\prod_{i=1}^{n-1} \de{u_i} \right) \ \omega^{n-1}, }
\ee
where $\omega$ was defined in \eqref{eqn:defom}. In particular for $n=2$, where $u=\frac{\kappa}{r}=\kappa\rho$
and $\omega=\frac{\kappa}{2} \sin\theta\, \de\theta \,\de\phi$, this formula boils down to 
\be
r^2 \de r \sin\theta\,\de\theta\, \de\phi \times
-\frac{\kappa^2}{4r^4} =  \frac12 \kappa\de\rho\, \times \frac12 \kappa \sin\theta\, \de\theta \,\de\phi\ ,
\ee
see Appendix~\ref{sec:proof-master-identity} for a proof for arbitrary $n$. 

\medskip

Inserting the identity \eqref{eqn:master-formula} in \eqref{eqn:genInRed}, we can trade the integral over 
$\va{x}_i$ into an integral over $\cM_n(\{\gamma_i,u_i\})$ at fixed values of $u_i$, further integrated
over $\IR^{n-1}$. The integral over $\cM_n(\{\gamma_i,u_i\})$ produces the symplectic volume \eqref{eqn:defgclass}, leading to 
\bea
 \label{eqn:integral-before-coord-change}
\cI_n(\{\gamma_i,c_i\},\beta) &=& 
\left( \frac{\beta}{4\pi} \right)^{n-1}
\int \prod_{i=1}^{n-1} 
 \de u_i\,  \de D_i\, 
\Vol(\{\gamma_i,u_i\})\, 
 e^{-\frac{\beta}{2} \left( \sum_{i=1}^{n-1} (c_i-u_i) D_i + \sum_{i,j=1}^{n-1}  \sfM_{ij}  D_i D_j\right)}
\\
&=&  \left( \frac{\beta}{4\pi} \right)^{n-1} \left(
\det \frac{\beta \sfM}{2\pi}
\right)^{-1/2} 
\int \prod_{i=1}^{n-1} 
 \de u_i\,  
\Vol(\{\gamma_i,u_i\})\, 
e^{-\frac{\beta}{8} \sum_{i,j=1}^{n-1} (u_i-c_i) \sfM^{-1}_{ij} 
(u_j-c_j)}\nn,
\eea
where in the second line we integrated over $D_i\in\IR^{n-1}-\I\epsilon$. It is convenient to change variables 
\be
u_i \to z_i = \sqrt\frac{\beta}{8\pi} u_j \sfE^{-1}_{j i},
\ee
where $\sfE$ is defined as a Cholesky-type factor for $\sfM$ through the relation $\sfE^T \sfE = \sfM$. Since there is an $\mathrm{O}(n-1)$ arbitrariness in $\sfE$, we choose it
for definiteness so that it is lower-triangular with a positive diagonal. The transformation above makes the quadratic form inside the exponent in \eqref{eqn:integral-before-coord-change} diagonal,
\be
\label{eqn:defxi}
  -\frac{\beta}{8} \sum_{i, j = 1}^{n-1} (u_i - c_i) \, \sfM^{-1}_{i j} \, (u_j - c_j) =  - \pi \sum_{i = 1}^{n-1} (z_i - \xi_i)^2, \quad \xi_i = \sqrt\frac{\beta}{8 \pi} c_i E^{-1}_{j i}.
\ee
Further, the measure $\prod_{i=1}^{n-1} \dd{u_i}$ becomes
\be
  \prod_{i=1}^{n-1} \dd{u_i} = \det(\sqrt\frac{8\pi}{\beta} \sfE^T) \prod_{i=1}^{n-1} \dd{z_i} = \qty(\frac{8\pi}{\beta})^\frac{n-1}{2} \sqrt{\det\sfM} \prod_{i=1}^{n-1} \dd{z_i},
\ee
such that the factor in front of the integral over $z_i$ becomes unity,
\be
  \qty(\frac{\beta}{4\pi})^{n-1} \qty(\det \frac{\beta \sfM}{2\pi})^{-1/2} \qty(\frac{8\pi}{\beta})^\frac{n-1}{2} \sqrt{\det\sfM} = 1,
\ee
justifying {\it a posteriori} the choice of normalization of the integral in \eqref{eqn:genIn}. The remaining integral can be written as
\be\label{eqn:genIny-final-integral}
 \boxed{  \cI_n(\{\gamma_i,c_i\},\beta) = \int \prod_{i=1}^{n-1} 
  \dd{z_i}  \,  
 e^{-\pi \sum_{i=1}^{n-1}(z_i - \xi_i)^2} \, \Vol(\{\gamma_i, \sfE^T \boldsymbol{z}\}). }
\ee
Similarly for the refined index, assuming that switching on the chemical potential $y$ has the effect of 
performing the same replacement \eqref{eqn:Aroofrep} on the r.h.s. of  \eqref{eqn:master-formula},  
 we obtain 
\be\label{eqn:genIny-final-integral-y}
 \boxed{  \cI_n(\{\gamma_i,c_i\},\beta,y) = \int \prod_{i=1}^{n-1} 
  \dd{z_i}  \,  
 e^{-\pi \sum_{i=1}^{n-1}(z_i - \xi_i)^2} \, \Ind_\mathrm{tree}(\{\gamma_i, \sfE^T \boldsymbol{z}\},y). }
\ee
We shall refer to the contribution of a particular permutation $\sigma$ in \eqref{eqn:gtree} as a `partial Witten index'.

In particular, for any term of the form $  \prod_{\alpha = 1}^{r} \sgn\qty(\sum_{i=1}^{n-1} c_i \sfH_{i \alpha})$ in  $\Vol(\{\gamma_i, c_i\})$ or $\Ind_\mathrm{tree}(\{\gamma_i,c_i\})$, 
the integral yields a term
\be
  \int \prod_{i=1}^{n-1} 
  \dd{z_i}  \,  
 e^{-\pi \sum_{i=1}^{n-1}(z_i - \xi_i)^2} \, \prod_{\alpha = 1}^{r} \sgn\qty(\sum_{i, j=1}^{n-1} z_j \sfE_{j i} \sfH_{i \alpha}),
\ee
which is clearly a generalized error function of depth $r$. To obtain its explicit form, we have to rotate the $z_i$'s such that the $r$ linear forms $\sum_{i, j=1}^{n-1} z_j \sfE_{j i} \sfH_{i \alpha}$ labeled by $\alpha$ only involve $z_1$ through $z_r$. This ensures that the integration over the remaining $z_{r+1},\dots z_{n-1}$ will be trivial and yield $1$. This rotation can be done by an orthogonalization procedure for the columns of $\sfE \sfH$. The result is a certain lower-triangular $r \times r$ matrix $\sfL$ and a row-orthonormal $r \times (n-1)$ matrix $\mathcal{O}$ such that $\sfE \sfH = \mathcal{O}^T \sfL$. Noting that the generalized error functions are invariant under arbitrary orthogonal transformations, and recalling the definition of $\xi$ in \eqref{eqn:defxi}, the integral can finally be written as
\be
  \boxed{\int \prod_{i=1}^{n-1} 
  \dd{z_i}  \,  
 e^{-\pi \sum_{i=1}^{n-1}(z_i - \xi_i)^2} \, \prod_{\alpha = 1}^{r} \sgn\qty(\sum_{i, j=1}^{n-1} z_j \sfE_{j i} \sfH_{i \alpha}) = E_r\qty(\sfL, \sqrt\frac{\beta}{8\pi} \, \mathcal{O} \, \sfE^{-T} c). }
\ee
Since $\mathcal{O}^T \sfL = \sfE \sfH$, the result can also be written as
\be
  E_r\qty(\sfL, \sqrt\frac{\beta}{8\pi} \, \mathcal{O} \, \sfE^{-T} c) = E_r\qty(\tilde{\sfL}, \sqrt\frac{\beta}{8\pi} \, \sfL^{-T} \, (\sfH^T c)),
\ee
where $\tilde{\sfL}$ is a lower triangular matrix with $1$s on the diagonal obtained by scaling the columns of $\sfL$ by the corresponding diagonal element (which we are allowed to do because of the scaling properties of the generalized error functions given in Section \ref{sec:gen-erf}). Thus, for any number $n$ of centers, the refined Witten index can be expressed as a sum of generalized error functions of depth $\leq n-1$, weighted by products of sign functions of the $\gamma_{ij}$'s and by 
factors of $y^{\gamma_{ij}}$.

\subsubsection{Explicit results for up to 4 centers}
\label{sec:results-four-centers}
We can now compute the Witten index for up to 4 centers using the prescription just outlined. 
The starting point is the formula for the partial tree index, computed by applying the recursion relation from Section \ref{sec:phase-space} (or by reading off the result from~\cite[Section C.1]{Alexandrov:2018iao}): 
\be
\label{eqn:F-tr-2}
  F_{\mathrm{tree}, 2} = \frac{1}{2} \qty\Big[\sgn(c_1) + \sgn(\gamma_{12})],
\ee
\be
\label{eqn:F-tr-3}
  \begin{aligned}
    F_{\text{tree},3} &= \frac{1}{4} \left[ 
      \left( \sgn(c_1) + \sgn(\gamma_{12}) \right)
      \left( \sgn(c_1 + c_2) + \sgn(\gamma_{23}) \right) \right. \\
      &\quad \left. - \left( \sgn(\gamma_{2+3, 1}) + \sgn(\gamma_{12}) \right)
      \left( \sgn(\gamma_{3, 1+2}) + \sgn(\gamma_{23}) \right)
    \right],
  \end{aligned}
\ee
\be
\label{eqn:F-tr-4}
  \begin{aligned}
    F_{\text{tree},4} = \frac{1}{8} \Big[ \,
    & (\sgn(c_1) + \sgn(\gamma_{12}))
      (\sgn(c_1+c_2) + \sgn(\gamma_{23}))
      (\sgn(c_1+c_2+c_3) + \sgn(\gamma_{34})) \\
    & - (\sgn(\gamma_{2+3+4, 1}) + \sgn(\gamma_{12}))
        (\sgn(\gamma_{3+4, 1+2}) + \sgn(\gamma_{23}))
        (\sgn(\gamma_{4, 1+2+3}) + \sgn(\gamma_{34})) \\
    & - (\sgn(c_1) - \sgn(\gamma_{2+3+4, 1}))
        (\sgn(\gamma_{3+4, 2}) + \sgn(\gamma_{23}))
        (\sgn(\gamma_{4,2+3}) + \sgn(\gamma_{34})) \\
    & - (\sgn(c_1+c_2+c_3) - \sgn(\gamma_{4, 1+2+3}))
        (\sgn(\gamma_{2+3, 1}) + \sgn(\gamma_{12}))
        (\sgn(\gamma_{3, 1+2}) + \sgn(\gamma_{23})) \, \Big],
  \end{aligned}
\ee
where we recall that the FI parameters $c_i$ are related to the central charges of the constituents via \eqref{eqn:FI-BPS}.
Using these partial tree indices as the starting point, one can compute the index \eqref{eqn:genIny}. We already know the result for two centers but now we can see it even more explicitly: inserting \eqref{eqn:F-tr-2} into our general formula \eqref{eqn:genIny-final-integral}, and noting that $\sfE = \sqrt{m} = \sqrt\frac{m_1 m_2}{m_1 + m_2}$ for two centers, we see that the effect of the integration over $u$ is to replace
\be
  \sgn(c_1) \to E_1\qty(\sqrt\frac{\beta}{8 \pi m} c_1),
\ee
leading to the partial Witten index
\be
\mathcal{J}_2 = \frac{1}{2} \qty\Big[\sgn(\gamma_{12}) + E_1\qty(\sqrt\frac{\beta}{8 \pi m} c_1) ] .
\ee
Using the fact that $\mathcal{J}_1$ is odd under the permutation $\gamma_1\leftrightarrow \gamma_2$, the sum over permutations gives the  refined index for two centers,
\be\label{eqn:two-centers-result}
  \mathcal{I}_{2}(\{\gamma_1 ,\gamma_2, c_1\}, \beta, y) = -\frac{(-1)^{\gamma_{1 2}}}{2} \, \qty(E_1\qty(\sqrt\frac{\beta}{8 \pi m} c_1) + \sgn(\gamma_{1 2})) \, \qty(\frac{y^{\gamma_{12}} - y^{-\gamma_{12}}}{y - y^{-1}}),
\ee
which agrees with \eqref{eqn:two-body-result-refined} up to an overall sign.
In the presence of a non-vanishing $\eta$ parameter, we expect that 
the argument of the $E_1$ will be shifted in opposite directions for the two permutations, such that the dependence in $y$ no longer factorizes, leading to the additional Gaussian term in the unrefined limit $y\to 1$. At the attractor point
$c_1=-2\lambda \gamma_{12}$, the partial Witten index simplifies to 
\be
\mathcal{J}_2^* = - \frac{1}{2} M_1\qty(\sqrt\frac{\lambda^2\tau_2}{m} \gamma_{12}),
\ee
where we have set $\beta=2\pi\tau_2$.

\subsection*{Three Centers}

For three centers, starting with the partial tree index \eqref{eqn:F-tr-3},
the $u$–integration implements the following replacements: \footnote{To avoid cluttering, we set $\beta=8\pi$. The parameter $\beta$ can be easily reinstated by rescaling $m\mapsto 8\pi m/\beta$ in the arguments $x_i$ of $E_n$, keeping the parameters $\cM$ fixed.}
\be
  \begin{aligned}
    \sgn(c_1)\,\sgn(c_1 + c_2)
      &\;\longrightarrow\;
        E_{2}\!\Bigl(
          \sqrt{\frac{m_{1}m_{3}}{m_{2}\,(m_{1}+m_{2}+m_{3})}};\,
          \frac{\,c_{1}m_{2}-c_{2}m_{1}\,}
               {\sqrt{m_{1}m_{2}(m_{1}+m_{2})}},\,
          \frac{c_{1}+c_{2}}{\sqrt{m_{1+2,3}}}
        \Bigr),\\[6pt]
    \sgn(c_1 + c_2)
      &\;\longrightarrow\;
        E_{1}\!\Bigl(\tfrac{c_{1}+c_{2}}{\sqrt{m_{1+2,3}}}\Bigr),\\[6pt]
    \sgn(c_1)
      &\;\longrightarrow\;
        E_{1}\!\Bigl(\tfrac{c_{1}}{\sqrt{m_{1,2+3}}}\Bigr).
  \end{aligned}
\ee
The contribution from the trivial permutation leads to the following `partial Witten index',
\be\label{eqn:three-centers-witten-index}
\begin{split}
\mathcal{J}_{3}
&= \frac{1}{4}\Bigl[
  E_{2}\!\Bigl(
    \sqrt{\frac{m_{1}m_{3}}{m_{2}(m_{1}+m_{2}+m_{3})}};\,
    \frac{c_{2}m_{1}-c_{1}m_{2}}{\sqrt{m_{1}m_{2}(m_{1}+m_{2})}},\,
    \frac{c_{3}}{\sqrt{m_{1+2,3}}}
  \Bigr)
  -\,\sgn(\gamma_{1,2+3})\,\sgn(\gamma_{1+2,3})\\[4pt]
&\quad
  -\,\bigl[E_{1}\!\bigl(\tfrac{c_{3}}{\sqrt{m_{1+2,3}}}\bigr)
          -\sgn(\gamma_{1+2,3})\bigr]\,
     \sgn(\gamma_{12})
  +\,\bigl[E_{1}\!\bigl(\tfrac{c_{1}}{\sqrt{m_{1,2+3}}}\bigr)
          -\sgn(\gamma_{2+3,1})\bigr]\,
     \sgn(\gamma_{23})
\Bigr].
\end{split}
\ee
where $m_{1+2,3}=\frac{m_3(m_1+m_2)}{(m_1+m_2+m_3)}$ denotes the reduced mass for the two-body problem $(\gamma_1+\gamma_2,\gamma_3)$, and similarly for $m_{1,2+3}$. Similarly, we denote $\gamma_{1+2,3}=\langle \gamma_1+\gamma_2,\gamma_3\rangle = \gamma_{12}+\gamma_{13}$.
Multiplying by $y^{\sum_{i<j}\gamma_{ij}}/(y-1/y)^2$ and summing over all permutations (for vanishing $\eta$), we can recombine the terms as follows:
\be
\begin{split}
\mathcal{I}_{3}
&= -\frac{1}{4}\Bigl[
  E_{2}\!\Bigl(
    \sqrt{\tfrac{m_{2}m_{3}}{m_{1}(m_{1}+m_{2}+m_{3})}};\,
    \tfrac{-\,c_{2}m_{1}+c_{1}m_{2}}{\sqrt{m_{1}m_{2}(m_{1}+m_{2})}},\,
    \tfrac{c_{3}}{\sqrt{m_{1+2,3}}}
  \Bigr)\,
  \kappa(\gamma_{13})\,\kappa(\gamma_{1+3,2})\\[4pt]
&\quad
  -\,E_{2}\!\Bigl(
    \sqrt{\tfrac{m_{1}m_{3}}{m_{2}(m_{1}+m_{2}+m_{3})}};\,
    \tfrac{c_{2}m_{1}-c_{1}m_{2}}{\sqrt{m_{1}m_{2}(m_{1}+m_{2})}},\,
    \tfrac{c_{3}}{\sqrt{m_{1+2,3}}}
  \Bigr)\,
  \kappa(\gamma_{23})\,\kappa(\gamma_{2+3,1})\\[4pt]
&\quad
  -\,E_{1}\!\Bigl(\tfrac{c_{3}}{\sqrt{m_{1+2,3}}}\Bigr)\,
     \kappa(\gamma_{12})\,\kappa(\gamma_{1+2,3})\,\sgn(\gamma_{12})
  -\,E_{1}\!\Bigl(\tfrac{c_{2}}{\sqrt{m_{1+3,2}}}\Bigr)\,
     \kappa(\gamma_{13})\,\kappa(\gamma_{1+3,2})\,\sgn(\gamma_{13})\\[4pt]
&\quad
  -\,E_{1}\!\Bigl(\tfrac{c_{1}}{\sqrt{m_{2+3,1}}}\Bigr)\,
     \kappa(\gamma_{23})\,\kappa(\gamma_{2+3,1})\,\sgn(\gamma_{23})
  +\,\kappa(\gamma_{2+3,1})\,\kappa(\gamma_{23})\\[4pt]
&\quad
  +\,\kappa(\gamma_{13})\,\kappa(\gamma_{1+3,2})\,
     \sgn(\gamma_{13})\,\sgn(\gamma_{1+3,2})
  +\,\kappa(\gamma_{2+3,1})\,\kappa(\gamma_{23})\,
     \sgn(\gamma_{2+3,1})\,\sgn(\gamma_{1+3,2})\\[4pt]
&\quad
  +\,\kappa(\gamma_{2+3,1})\,\kappa(\gamma_{23})\,
     \sgn(\gamma_{2+3,1})\,\sgn(\gamma_{23})
  +\,\kappa(\gamma_{12})\,\kappa(\gamma_{1+2,3})\,
     \sgn(\gamma_{12})\,\sgn(\gamma_{1+2,3})\\[4pt]
&\quad
  -\,\kappa(\gamma_{12})\,\kappa(\gamma_{1+2,3})\,
     \sgn(\gamma_{1+3,2})\,\sgn(\gamma_{1+2,3})
\Bigr].
\end{split}
\ee
where we denoted by $\kappa(p)=(-1)^p \frac{y^p-y^{-p}}{y-1/y}$ the character of a spin $(|p|-1)/2$ representation of $SO(3)$, and used the functional equation \eqref{eqn:E2trial} to eliminate one of the three $E_2$ functions. 
At the attractor point
$c_1 = 2\lambda \gamma_{2+3, 1}$, $c_2 = 2\lambda \gamma_{1+3, 2}$, 
using the identities \eqref{eqn:E1fromM}--\eqref{eqn:E2fromM},
the partial Witten index reduces to 
\be
  \begin{aligned}
   \mathcal{J}_3^* &= \frac{1}{4} \Biggl( M_2\left(\sqrt{\frac{m_1 m_3}{m_2 \left(m_1+m_2+m_3\right)}};-\frac{\sqrt{\lambda^2\tau_2} \left(m_2 \gamma _{1, 2+3} - m_1 \gamma _{2, 1+3}\right)}{\sqrt{m_1 m_2 \left(m_1+m_2\right)}},-\frac{\sqrt{\lambda^2\tau_2} \gamma _{1+2, 3}}{\sqrt{m_{1+2, 3}}}\right) \\
    &\qquad + M_1\left(\frac{\sqrt{\lambda^2\tau_2} \gamma _{1+2, 3}}{\sqrt{m_{1+2, 3}}}\right) \left(\sgn\left(m_2 \gamma _{1,2+3}-m_1 \gamma _{2, 1+3}\right)-\sgn\left(\gamma _{12}\right)\right) \\
    &\qquad + M_1\left(\frac{\sqrt{\lambda^2\tau_2} \gamma _{1, 2+3}}{\sqrt{m_{1, 2+3}}}\right) \left(\sgn\left(m_2 \gamma _{1+2,3}-m_3 \gamma _{1+3, 2}\right)-\sgn\left(\gamma _{2 3}\right)\right) \Biggr).
  \end{aligned}
\ee
The terms proportional to $M_2$ are identified as the contribution to the Witten index from the continuum of scattering states for the 3-body supersymmetric quantum mechanics.
The terms proportional to $M_1$ arise instead from scattering states in the 2-body problem, where one of the bodies is a two-particle bound state. In the special case where all D4-brane charges are collinear, we observe that the coefficients of $M_1$ vanish,\footnote{For example, setting $p_i=N_i p_0$ with $N_i>0$ and $(p_0)^3>0$, 
$m_2 \gamma _{1,2+3}-m_1 \gamma _{2, 1+3}=(p_0^3) (N_1+N_2+N_3)  \gamma_{12}$.
} \ie two-particle bound states are absent in the attractor chamber, leaving only the
contribution from the 3-body continuum of scattering states.

\subsection*{Four Centers}

For four centers, it is convenient to introduce
\be
  \begin{aligned}
    \alpha &= \sqrt{\frac{m_{1}m_{3}}{m_{2}(m_{1}+m_{2}+m_{3})}},\quad
    \beta  = \sqrt{\frac{m_{1}(m_{1}+m_{2})m_{4}}
                       {m_{2}(m_{1}+m_{2}+m_{3})(m_{1}+m_{2}+m_{3}+m_{4})}},\\[4pt]
    \gamma &= \sqrt{\frac{(m_{1}+m_{2})m_{4}}
                       {m_{3}(m_{1}+m_{2}+m_{3}+m_{4})}}.
  \end{aligned}
\ee
We then define the building blocks
\begin{align}\label{eqn:Phi-building-blocks}
  \Phi_{3}&= E_{3}\!\Bigl(
    \alpha,\;\beta,\;\gamma;\;
    \tfrac{c_{1}m_{2}-c_{2}m_{1}}{\sqrt{m_{1}m_{2}(m_{1}+m_{2})}},\;
    \tfrac{(c_{1}+c_{2})m_{3}-c_{3}(m_{1}+m_{2})}
         {\sqrt{(m_{1}+m_{2})m_{3}(m_{1}+m_{2}+m_{3})}},\;
    \tfrac{-\,c_{4}}{\sqrt{m_{1+2+3,4}}}
  \Bigr),\nonumber\\[4pt]
  \Phi_{2}^{(12)}&= E_{2}\!\Bigl(
    \sqrt{\tfrac{(m_{1}+m_{2})m_{4}}
                  {m_{3}(m_{1}+m_{2}+m_{3}+m_{4})}};\;
    \tfrac{c_{3}(m_{1}+m_{2})-(c_{1}+c_{2})m_{3}}
         {\sqrt{(m_{1}+m_{2})m_{3}(m_{1}+m_{2}+m_{3})}},\;
    \tfrac{\,c_{4}}{\sqrt{m_{1+2+3,4}}}
  \Bigr),\nonumber\\[4pt]
  \Phi_{2}^{(23)}&= E_{2}\!\Bigl(
    \sqrt{\frac{m_{1}m_{4}}{(m_{2}+m_{3})(m_{1}+m_{2}+m_{3}+m_{4})}};\;
    \frac{(c_{2}+c_{3})m_{1}-c_{1}(m_{2}+m_{3})}
         {\sqrt{m_{1}(m_{2}+m_{3})(m_{1}+m_{2}+m_{3})}},\;
    \tfrac{\,c_{4}}{\sqrt{m_{1+2+3,4}}}
  \Bigr),\nonumber\\[4pt]
  \Phi_{2}^{(34)}&= E_{2}\!\Bigl(
    \sqrt{\frac{m_{1}(m_{3}+m_{4})}
               {m_{2}(m_{1}+m_{2}+m_{3}+m_{4})}};\;
    \frac{c_{1}m_{2}-c_{2}m_{1}}{\sqrt{m_{1}m_{2}(m_{1}+m_{2})}},\;
    \frac{c_{1}+c_{2}}{\sqrt{m_{1+2,3+4}}}
  \Bigr).
\end{align}
The partial Witten index then takes the form
\be\label{eqn:partial-witten-4}
  \begin{split}
    \mathcal{J}_4 = \frac{1}{8} \Big[
    &\Phi_{3} \;
    +\;\Phi_{2}^{(12)} \, \sgn(\gamma_{1 2})
    \;+\;\Phi_{2}^{(23)} \, \sgn(\gamma_{2 3})
    \;+\;\Phi_{2}^{(34)} \, \sgn(\gamma_{3 4}) \\
    &+\;E_1\qty(\frac{c_1}{\sqrt{m_{1,2+3+4}}}) \;\,
      \begin{split}
        \bigl[ &\sgn(\gamma_{2,3+4})\,\sgn(\gamma_{3 4})
        \;+\;\sgn(\gamma_{2 3})\,\sgn(\gamma_{2+3,4}) \\
        &\;-\;\sgn(\gamma_{2,3+4})\,\sgn(\gamma_{2+3,4}) \bigr]
      \end{split} \\[6pt]
    &+\; E_1\qty(\frac{c_1 + c_2}{\sqrt{m_{1+2, 3+4}}}) \;\,\sgn(\gamma_{1 2})\,\sgn(\gamma_{3 4}) \\[6pt]
    &+\; E_1\qty(\frac{c_1 + c_2 + c_3}{\sqrt{m_{1+2+3, 4}}}) \;\, 
    \begin{split}
      \bigl[ &\sgn(\gamma_{1,2+3})\,\sgn(\gamma_{2 3})
      \;+\;\sgn(\gamma_{1 2})\,\sgn(\gamma_{1+2,3}) \\
      &\;-\;\sgn(\gamma_{1,2+3})\,\sgn(\gamma_{1+2,3}) \bigr]
    \end{split} \\[6pt]
    &+\;\bigl[
    \sgn(\gamma_{1,2+3+4})\,\sgn(\gamma_{2,3+4})\,\sgn(\gamma_{3 4})
    \;+\;\sgn(\gamma_{1 2})\,\sgn(\gamma_{1+2,3+4})\,\sgn(\gamma_{3 4})\\
    &\qquad\;-\;\sgn(\gamma_{1,2+3+4})\,\sgn(\gamma_{1+2,3+4})\,\sgn(\gamma_{3 4})
    \;+\;\sgn(\gamma_{1,2+3+4})\,\sgn(\gamma_{2 3})\,\sgn(\gamma_{2,3+4})\\
    &\qquad\;-\;\sgn(\gamma_{1,2+3+4})\,\sgn(\gamma_{2,3+4})\,\sgn(\gamma_{2+3,4})
    \;+\;\sgn(\gamma_{1,2+3})\,\sgn(\gamma_{2 3})\,\sgn(\gamma_{1+2+3,4})\\
    &\qquad\;-\;\sgn(\gamma_{1,2+3+4})\,\sgn(\gamma_{2 3})\,\sgn(\gamma_{1+2+3,4})
    \;+\;\sgn(\gamma_{1 2})\,\sgn(\gamma_{1+2,3})\,\sgn(\gamma_{1+2+3,4})\\
    &\qquad\;-\;\sgn(\gamma_{1,2+3})\,\sgn(\gamma_{1+2,3})\,\sgn(\gamma_{1+2+3,4})
    \;-\;\sgn(\gamma_{1 2})\,\sgn(\gamma_{1+2,3+4})\,\sgn(\gamma_{1+2+3,4})\\
    &\qquad\;+\;\sgn(\gamma_{1,2+3+4})\,\sgn(\gamma_{1+2,3+4})\,\sgn(\gamma_{1+2+3,4})
    \bigr] \Big].
  \end{split}
\ee
The full index is obtained by multiplying by $y^{\sum_{i<j}\gamma_{ij}}/(y-1/y)^3$ and summing over all permutations. 
At the attractor point, using \eqref{eqn:E1fromM}--\eqref{eqn:E3fromM} this reduces to
\begin{align}
\mathcal{J}_4^* &= \frac{1}{8}\Bigl\{
  H_{3}
  + H_{2}^{(12)}\Bigl[\sgn(\gamma_{12})
    - \sgn\bigl(\gamma_{1+3+4,2}\,m_{1}
      - \gamma_{2+3+4,1}\,m_{2}\bigr)\Bigr]\notag\\
&\quad
  + H_{2}^{(23)}\Bigl[\sgn(\gamma_{23})
    - \sgn\bigl(\gamma_{1+2+4,3}\,m_{2}
      - \gamma_{1+3+4,2}\,m_{3}\bigr)\Bigr]\notag\\
&\quad
  + H_{2}^{(34)}\Bigl[\sgn(\gamma_{34})
    - \sgn\bigl(\gamma_{1+2+3,4}\,m_{3}
      - \gamma_{1+2+4,3}\,m_{4}\bigr)\Bigr]\notag\\
&\quad
  + M_{1}\!\Bigl(\tfrac{c_{1}}{\sqrt{m_{1,2+3+4}}}\Bigr)\Bigl[
      \sgn(\gamma_{2,3+4})
        \bigl(\sgn(\gamma_{34})
        - \sgn(\gamma_{2+3,4})\bigr)\notag\\
&\qquad\quad
    - \sgn\bigl(\gamma_{1,2+3+4}\,m_{2}
        + \gamma_{2,1+3+4}\,m_{2+3+4}\bigr)
      \bigl(\sgn(\gamma_{34})
        - \sgn\bigl(\gamma_{1,2+3+4}\,m_{2+3}
            - \gamma_{1+4,2+3}\,m_{2+3+4}\bigr)\bigr)\notag\\
&\qquad\quad
    + \sgn(\gamma_{23})
      \bigl(\sgn(\gamma_{2,3+4})
        - \sgn\bigl(\gamma_{1,2+3+4}\,m_{2+3}
            - \gamma_{1+4,2+3}\,m_{2+3+4}\bigr)\bigr)
  \Bigr]\notag\\
&\quad
  + M_{1}\!\Bigl(\tfrac{c_{1}+c_{2}}{\sqrt{m_{1+2,3+4}}}\Bigr)\Bigl[
      \sgn(\gamma_{12})\,\sgn(\gamma_{34})
    - \sgn(\gamma_{34})\,
      \sgn\bigl(-\gamma_{2,3+4}\,m_{1}
        + \gamma_{1,3+4}\,m_{2}
        + \gamma_{12}(m_{1}+m_{2})\bigr)\notag\\
&\qquad\quad
    - \sgn(\gamma_{12})\,
      \sgn\bigl(\gamma_{1+2+3,4}\,m_{3}
        - \gamma_{1+2,3}\,m_{4}
        + \gamma_{34}\,m_{4}\bigr)\notag\\
&\qquad\quad
    + \sgn\bigl(-\gamma_{2,3+4}\,m_{1}
        + \gamma_{1,3+4}\,m_{2}
        + \gamma_{12}(m_{1}+m_{2})\bigr)\,
      \sgn\bigl(\gamma_{1+2+3,4}\,m_{3}
        - \gamma_{1+2,3}\,m_{4}
        + \gamma_{34}\,m_{4}\bigr)
  \Bigr]\notag\\
&\quad
  + M_{1}\!\Bigl(\tfrac{c_{1}+c_{2}+c_{3}}{\sqrt{m_{1+2+3,4}}}\Bigr)\Bigl[
      \sgn(\gamma_{1,2+3})\,\sgn(\gamma_{23})
    + \sgn(\gamma_{12})\,\sgn(\gamma_{1+2,3})\notag\\
&\qquad\quad
    - \sgn(\gamma_{1,2+3})\,\sgn(\gamma_{1+2,3})
    + \sgn(\gamma_{12})\,
      \sgn\bigl(-2(\gamma_{1+2,3}-\gamma_{34})(m_{1}+m_{2})
        -2\,\gamma_{1,3+4}\,m_{3}\bigr)\notag\\
&\qquad\quad
    + \sgn(\gamma_{23})\,
      \sgn\bigl(-2(\gamma_{1,2+3}-\gamma_{2+3,4})\,m_{1}
        -2\,\gamma_{1,2+3+4}(m_{2}+m_{3})\bigr)\notag\\
&\qquad\quad
    - \sgn\bigl(-2(\gamma_{1+2,3}-\gamma_{34})(m_{1}+m_{2})
        -2\,\gamma_{1,3+4}\,m_{3}\bigr)\, \notag\\
      &\qquad\quad\sgn\bigl(-\gamma_{2+3,4}\,m_{1}
        - \gamma_{34}\,m_{1}
        + \gamma_{14}(m_{2}+m_{3})
        + \gamma_{12}(m_{1}+m_{2}+m_{3})
        + \gamma_{13}(m_{1}+m_{2}+m_{3})\bigr)
  \Bigr]
\Bigr\}.
\end{align}
where as in \cref{eqn:Phi-building-blocks}, we define
\be
\begin{aligned}
  H_3 &\;=\; M_3\Big(\alpha,\; \beta,\; \gamma;\; 
    \tfrac{2\;\gamma_{2+3+4,1}\,m_2 \;-\; 2\;\gamma_{1+3+4,2}\,m_1}
    {\sqrt{m_1\,m_2\,(m_1 + m_2)}}, \;
    \tfrac{2\;\gamma_{3+4,1+2}\,m_3\;-\;2\;\gamma_{1+2+4,3}\,(m_1 + m_2)}
    {\sqrt{(m_1 + m_2)\,m_3\,(m_1 + m_2 + m_3)}}, \,
    \tfrac{-2\;\gamma_{1+2+3,4}}{\sqrt{m_{1+2+3, 4}}}
  \Big), \\
  H_2^{(12)} &\;=\; M_2\Big(
    \sqrt{\tfrac{(m_1 + m_2)\,m_4}{m_3\,(m_1 + m_2 + m_3 + m_4)}}; \;
    \tfrac{2\;\gamma_{3+4,1+2}\,m_3\;-\;2\;\gamma_{1+2+4,3}\,(m_1 + m_2)}
    {\sqrt{(m_1 + m_2)\,m_3\,(m_1 + m_2 + m_3)}}, \;
    \tfrac{-2\;\gamma_{1+2+3,4}}{\sqrt{m_{1+2+3, 4}}}
  \Big), \\
  H_2^{(23)} &\;=\; M_2\Big(
    \sqrt{\tfrac{m_1\,m_4}{(m_2 + m_3)\,(m_1 + m_2 + m_3 + m_4)}}; \;
    -2\tfrac{\gamma_{1+4,2+3}\,m_1 \;+\;\gamma_{1,2+3+4}\,(m_2 + m_3)}
    {\sqrt{m_1\,(m_2 + m_3)\,(m_1 + m_2 + m_3)}}, \\
    &\qquad\qquad -2\tfrac{\gamma_{1+4,2+3}\,m_1 \;+\;\gamma_{1,2+3+4}\,(m_2 + m_3)} {\sqrt{m_1\,(m_2 + m_3)\,(m_1 + m_2 + m_3)}}
  \Big), \\
  H_2^{(34)} &\;=\; M_2\Big(
    \sqrt{\tfrac{m_1\,(m_3 + m_4)}{m_2\,(m_1 + m_2 + m_3 + m_4)}}; \;
    -2\tfrac{\gamma_{1+3+4,2}\,m_1 \;+\; \gamma_{1,2+3+4}\,m_2}
    {\sqrt{m_1\,m_2\,(m_1 + m_2)}}, \;
    -2\,\gamma_{1+2,3+4}\,
    \sqrt{\tfrac{m_1 + m_2 + m_3 + m_4}{(m_1 + m_2)\,(m_3 + m_4)}}
  \Big).
\end{aligned}
\ee
The coefficient of $M_3$ is identified as the contribution from the continuum of scattering states for the 4-body supersymmetric quantum mechanics.
The terms proportional to $M_2$ arise instead from scattering states in the 3-body problem, where one of the bodies is a two-particle bound state, and those proportional to $M_1$ arise from scattering states in the 2-body problem, where each of the two bodies is a two-particle bound state, or one is a three-particle bound state and the other is the remaining center. Again, in cases where magnetic charges are collinear, we observe that the coefficients of $M_2$ and $M_1$ vanish, leaving only the contribution from 
the continuum of 4-body scattering states.

\section{Comparison with modular predictions}
\label{sec:comparison}

In this section, we compare the results of the localization computation in Section \ref{sec:computing-witten-index} with the prescription for the 
modular completion reviewed in Section \ref{sec:mock-modularity}. We first consider
the special case of local $\mathbb{P}^2$,  which was treated in detail in \cite[Section  4]{Alexandrov:2019rth}. Returning to the case of a generic CY threefold, 
we then compare the Witten index of the 
supersymmetric quantum mechanics with the instanton generating potential.

\subsection{Vafa-Witten invariants on \texorpdfstring{$\IP^2$}{P²}}

As explained in \cite[Section  4]{Alexandrov:2019rth}, the non-compact CY threefold 
$X=K_{\IP^2}=\mathcal{O}_{\mathbb{P}^2}(-3)$ (\ie 
the total space of the canonical bundle over the complex projective plane)
arises in the large fiber limit of a smooth elliptic fibration $\cE\to \tilde X\to \IP^2$. The D4-D2-D0 indices counting D4-branes wrapped on $N$ times the section $S\simeq \IP^2$ coincide with rank $N$ Vafa-Witten (VW) invariants of $\IP^2$, which have been computed by various methods (see \cite{Beaujard:2020sgs} for a review). Even though the section $S$ itself is not an ample divisor, it was conjectured in \cite{Alexandrov:2019rth} that the modular anomaly equations continue to hold, predicting the modular completion of generating series of VW invariants at arbitrary rank. 
For rank 2 and 3, it was shown to reproduce the known modular completions obtained in~\cite{Bringmann:2010sd,Manschot:2017xcr}. 
The modular anomaly was subsequently derived in \cite{Dabholkar:2020fde} by studying the elliptic genus of the superconformal field theory obtained by reducing the (0,2) six-dimensional SCFT of type $A_{N-1}$ on $\IP^2$.
Our goal is to reproduce the modular completions for rank $N\leq 4$ from the Witten index of the supersymmetric quantum mechanics of $n$ dyons computed in the previous subsection. 

\medskip

Following \cite[Appendix F]{Alexandrov:2019rth}, we define the normalized generating series of rank $N$ refined Vafa-Witten invariants on $\IP^2$ as
\be
g_{N, \mu} (\tau, w) := \frac{
h_{N, \mu+\frac12N(N-1)}(\tau, w+\frac12)}
{\qty(h_{1, 0}(\tau, w+\frac12))^N}, \qquad h_{1, 0}(\tau, w) = \frac{\I}{\theta_1(\tau, 2 w)},
\ee
where $\mu$ denotes the D2-brane charge (or 't Hooft flux) along the hyperplane class of $\IP^2$, and the rank one result $h_{1, 0}(\tau, w)$ follows from G\"ottsche's formula \cite{Gottsche:1990}.
Similarly, we define the normalized completed series  $\widehat{g}_{N, \mu}$ as the ratio
$\widehat{h}_{N, \mu+\frac12N(N-1)}(\tau, w+\frac12)/(h_{1, 0}(\tau, w+\frac12))^N$. 

In order to compare to the explicit formulae for $\widehat{g}_{N, \mu}$
in \cite[Appendix F]{Alexandrov:2019rth}, we need to evaluate the masses $m_i$ and FI parameters $c_i$ for the relevant constituents. The total D4-brane charge $N[S]$ can a priori split into any number
$n\leq N$ of constituents with charge $N_i[S]$, such that $N=\sum_{i=1}^n N_i$. Correspondingly, the D2-brane charge $q[H]$ along the hyperplane class $H$ in $\IP^2$  splits into $q=\sum_{i=1}^n q_i$, such that each constituent carries
reduced charge $\check{\gamma_i}=(N_i,q_i)$. The DSZ pairing is given by 
\be
\langle \gamma_i, \gamma_j \rangle = 3 (N_i q_j - N_j q_i),
\ee
where the factor $3$ comes from the intersection $[S]\cap [H]=-3$. The mass and FI parameters of each constituent at the large volume attractor point
are given by 
\be
c_i = 2\lambda\,\langle \gamma,\gamma_i\rangle
      = 6\lambda\,(N\,q_i - N_i\,q), \qquad
m_i = \tfrac12\,\lambda^2\,(p_i\,p^2)
      = \tfrac{9}{2}\,\lambda^2\,N^2\,N_i,
\ee
where we used the fact that $[S]^3=c_1(\IP^2)^2=9$.

\subsection*{Rank 2}

For $N=2$, the only contributing partition has two constituents, with $N_1=N_2=1$. The modular completion is given by  \cite[(F.4)]{Alexandrov:2019rth}
\be\label{eqn:mod-completion-two-centers}
  \widehat{g}_{2,\mu} = g_{2,\mu} + \frac{1}{2} \sum_{\ell \in \mathbb{Z} + \mu/2} \qty[ E_1(\sqrt{\tau_2}(2\ell + 6\eta)) - \sgn\ell ] \, q^{-\ell^2} y^{6\ell}.
\ee
Upon identifying $\gamma_{1 2} = 6 \ell=3(q_2-q_1)$, we compute 
\be
c_1 = 6\lambda\,(q_1 - q_2)
      = -12\lambda\,\ell, \qquad
m_1 = m_2 
      = 18\lambda^2\,.
\ee
The result from our index computation \eqref{eqn:two-centers-result} thus reads (after identifying $\beta = 2\pi \tau_2$)
\be
  \mathcal{I}_{2} = \frac{(-1)^{\gamma_{1 2}}}{2} \, \qty(E_1\qty(\sqrt{\tau_2} 2 \ell) - \sgn(\ell)) \, y^{6 \ell}
  =  \frac{(-1)^{\gamma_{1 2}}}{2}  M_1\qty(\sqrt{\tau_2} 2 \ell),
\ee
which matches with \eqref{eqn:mod-completion-two-centers} up to an overall sign in the limit $\eta \to 0$.

\subsection*{Rank 3}

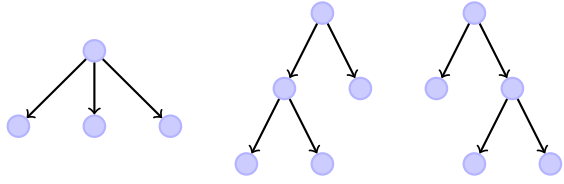
\begin{figure}[ht]
\begin{center}
\tikzstyle{place}=[circle,draw=blue!30,fill=blue!20,thick]
\begin{tikzpicture}[inner sep=1mm,scale=1,thick]
\begin{scope}
\node (a) at ( 0,0) [place] {};
\node (a1) at ( -1,-1) [place] {};
\node (a2)  at ( 0,-1) [place] {};
\node (a3)  at ( 1,-1) [place] {};
 \draw [->] (a) to node[auto] {} (a1);
 \draw [->] (a) to node[auto] {} (a2);
 \draw [->] (a) to node[auto] {} (a3);
 \end{scope}
 \begin{scope}[shift={(3,1/2)}]
 \node (a) at ( 0,0) [place] {};
\node (a1) at ( -1/2,-1) [place] {};
\node (a2)  at ( 1/2,-1) [place] {};
\node (a3)  at ( -1,-2) [place] {};
\node (a4)  at ( 0,-2) [place] {};
 \draw [->] (a) to node[auto] {} (a1);
 \draw [->] (a) to node[auto] {} (a2);
 \draw [->] (a1) to node[auto] {} (a3);
  \draw [->] (a1) to node[auto] {} (a4);
 \end{scope}
  \begin{scope}[shift={(5,1/2)}]
 \node (a) at ( 0,0) [place] {};
\node (a1) at ( -1/2,-1) [place] {};
\node (a2)  at ( 1/2,-1) [place] {};
\node (a3)  at ( 0,-2) [place] {};
\node (a4)  at ( 1,-2) [place] {};
 \draw [->] (a) to node[auto] {} (a1);
 \draw [->] (a) to node[auto] {} (a2);
 \draw [->] (a2) to node[auto] {} (a3);
  \draw [->] (a2) to node[auto] {} (a4);
 \end{scope}
\end{tikzpicture}
\end{center}
\caption{Schr\"oder trees contributing to the modular completion  for $N=3, N_i=1$
\label{fig:schroder-trees-3}}
\end{figure}

For $N=3$, two possible partitions contribute to the total result shown in \cite[(F.10)]{Alexandrov:2019rth}, namely $3=2+1$ and $3=1+1+1$. The contribution from the
partition matches the Witten index for 2-body quantum mechanics, exactly as in the rank 2 case above. We focus on the contribution of the second partition which arises from the 
three Schr\"oder trees shown in \cref{fig:schroder-trees-3}:
\be\label{eqn:mod-completion-g3}
\begin{aligned}
\Delta \widehat g_{3, \mu}
&= \frac{1}{4}
  \sum_{q_{1}+q_{2}+q_{3} = \mu + \tfrac{3}{2}}
  (-y)^{6(q_{3}-q_{1})}
  \exp\!\Bigl[-\tfrac{\tau}{6}\bigl((q_{2}-q_{1})^{2}+(q_{3}-q_{2})^{2}+(q_{3}-q_{1})^{2}\bigr)\Bigr] \\[6pt]
&\quad\times
  \Bigl[
    E_{2}\Bigl(\tfrac{1}{\sqrt{3}};\,\sqrt{\tau_{2}}(q_{2}-q_{1}+6\eta),\,
          \sqrt{\tfrac{\tau_{2}}{3}}(2q_{3}-q_{1}-q_{2}+18\eta)\Bigr) \\[-2pt]
&\qquad\quad
    -\;\sgn(q_{2}+q_{3}-2q_{1})\,\sgn(2q_{3}-q_{1}-q_{2})
    \;-\;\tfrac{1}{3}\,\delta_{q_{1}=q_{2}=q_{3}} \\[-2pt]
&\qquad\quad
    -\;\bigl(E_{1}(\sqrt{\tfrac{\tau_{2}}{3}}\,(2q_{3}-q_{1}-q_{2}+18\eta))
           -\sgn(2q_{3}-q_{1}-q_{2})\bigr)\,\sgn(q_{2}-q_{1}) \\[-2pt]
&\qquad\quad
    -\;\bigl(E_{1}(\sqrt{\tfrac{\tau_{2}}{3}}\,(q_{2}+q_{3}-2q_{1}+18\eta))
           -\sgn(q_{2}+q_{3}-2q_{1})\bigr)\,\sgn(q_{3}-q_{2})
  \Bigr].
\end{aligned}
\ee
One verifies that under the substitutions
\be
  c_i \;=\; 6 \lambda^2 \, (3\,q_i - \,(q_1+q_2+q_3)),
\qquad
  m_i \;=\;\tfrac{81}{2} \lambda^2,
\ee
the partial Witten index \eqref{eqn:three-centers-witten-index}
reduces exactly to the modular-completion \eqref{eqn:mod-completion-g3} above,  up to the $\eta$-shifts and the Kronecker-delta term 
$\tfrac13\,\delta_{q_1=q_2=q_3}$. In fact, up to the same terms, \eqref{eqn:mod-completion-g3} can be rewritten in terms of the $M_2$ function only, 
\be
\begin{aligned}
\Delta \widehat g_{3, \mu}
&= \frac{1}{4}
  \sum_{q_{1}+q_{2}+q_{3} = \mu + \tfrac{3}{2}}
  (-y)^{6(q_{3}-q_{1})}
  \exp\!\Bigl[-\tfrac{\tau}{6}\bigl((q_{2}-q_{1})^{2}+(q_{3}-q_{2})^{2}+(q_{3}-q_{1})^{2}\bigr)\Bigr] \\[6pt]
&\qquad\quad\times
  M_{2}\Bigl(\tfrac{1}{\sqrt{3}};\,\sqrt{\tau_{2}}(q_{2}-q_{1}+6\eta),\,
          \sqrt{\tfrac{\tau_{2}}{3}}(2q_{3}-q_{1}-q_{2}+18\eta)\Bigr).
\end{aligned}
\ee

\subsection*{Rank 4}

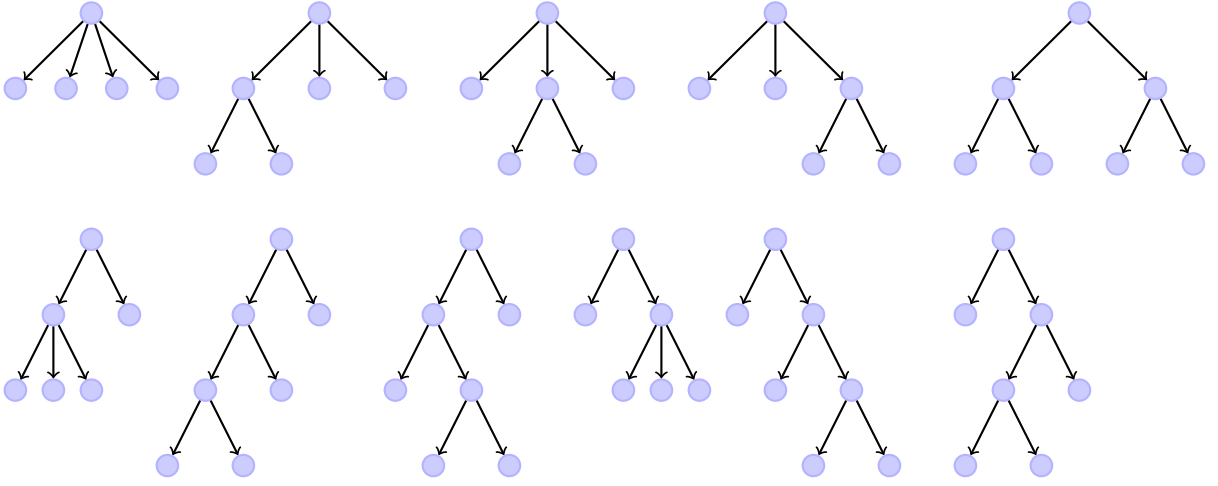
\begin{figure}[ht]
\begin{center}
\tikzstyle{place}=[circle,draw=blue!30,fill=blue!20,thick]
\begin{tikzpicture}[inner sep=1mm,scale=1,thick]
\begin{scope}[shift={(-1,0)}]
\node (a) at ( 0,0) [place] {};
\node (a1) at ( -1,-1) [place] {};
\node (a2)  at ( -1/3,-1) [place] {};
\node (a3)  at ( 1/3,-1) [place] {};
\node (a4)  at ( 1,-1) [place] {};
 \draw [->] (a) to node[auto] {} (a1);
 \draw [->] (a) to node[auto] {} (a2);
 \draw [->] (a) to node[auto] {} (a3);
  \draw [->] (a) to node[auto] {} (a4);
 \end{scope}
\begin{scope}[shift={(2,0)}]
\node (a) at ( 0,0) [place] {};
\node (a1) at ( -1,-1) [place] {};
\node (a2)  at ( 0,-1) [place] {};
\node (a3)  at ( 1,-1) [place] {};
\node (a4)  at ( -1/2,-2) [place] {};
\node (a5)  at ( -3/2,-2) [place] {};
 \draw [->] (a) to node[auto] {} (a1);
 \draw [->] (a) to node[auto] {} (a2);
 \draw [->] (a) to node[auto] {} (a3);
  \draw [->] (a1) to node[auto] {} (a4);
   \draw [->] (a1) to node[auto] {} (a5);
 \end{scope}
 \begin{scope}[shift={(5,0)}]
\node (a) at ( 0,0) [place] {};
\node (a1) at ( -1,-1) [place] {};
\node (a2)  at ( 0,-1) [place] {};
\node (a3)  at ( 1,-1) [place] {};
\node (a4)  at ( -1/2,-2) [place] {};
\node (a5)  at ( 1/2,-2) [place] {};
 \draw [->] (a) to node[auto] {} (a1);
 \draw [->] (a) to node[auto] {} (a2);
 \draw [->] (a) to node[auto] {} (a3);
  \draw [->] (a2) to node[auto] {} (a4);
   \draw [->] (a2) to node[auto] {} (a5);
 \end{scope}
 \begin{scope}[shift={(8,0)}]
\node (a) at ( 0,0) [place] {};
\node (a1) at ( -1,-1) [place] {};
\node (a2)  at ( 0,-1) [place] {};
\node (a3)  at ( 1,-1) [place] {};
\node (a4)  at ( 1/2,-2) [place] {};
\node (a5)  at ( 3/2,-2) [place] {};
 \draw [->] (a) to node[auto] {} (a1);
 \draw [->] (a) to node[auto] {} (a2);
 \draw [->] (a) to node[auto] {} (a3);
  \draw [->] (a3) to node[auto] {} (a4);
   \draw [->] (a3) to node[auto] {} (a5);
 \end{scope}
  \begin{scope}[shift={(12,0)}]
\node (a) at ( 0,0) [place] {};
\node (a1) at ( -1,-1) [place] {};
\node (a2)  at ( 1,-1) [place] {};
\node (a3)  at ( -3/2,-2) [place] {};
\node (a4)  at ( -1/2,-2) [place] {};
\node (a5)  at ( 1/2,-2) [place] {};
\node (a6)  at ( 3/2,-2) [place] {};
 \draw [->] (a) to node[auto] {} (a1);
 \draw [->] (a) to node[auto] {} (a2);
 \draw [->] (a1) to node[auto] {} (a3);
  \draw [->] (a1) to node[auto] {} (a4);
   \draw [->] (a2) to node[auto] {} (a5);
      \draw [->] (a2) to node[auto] {} (a6);
 \end{scope}
 \begin{scope}[shift={(-1,-3)}]
 \node (a) at ( 0,0) [place] {};
\node (a1) at ( -1/2,-1) [place] {};
\node (a2)  at ( 1/2,-1) [place] {};
\node (a3)  at ( -1,-2) [place] {};
\node (a4)  at ( -1/2,-2) [place] {};
\node (a5)  at ( 0,-2) [place] {};
\draw [->] (a) to node[auto] {} (a1);
\draw [->] (a) to node[auto] {} (a2);
\draw [->] (a1) to node[auto] {} (a3);
\draw [->] (a1) to node[auto] {} (a4);
\draw [->] (a1) to node[auto] {} (a5);
\end{scope}
  \begin{scope}[shift={(3/2,-3)}]
 \node (a) at ( 0,0) [place] {};
\node (a1) at ( -1/2,-1) [place] {};
\node (a2)  at ( 1/2,-1) [place] {};
\node (a3)  at ( -1,-2) [place] {};
\node (a4)  at ( 0,-2) [place] {};
\node (a5)  at ( -3/2,-3) [place] {};
\node (a6)  at ( -1/2,-3) [place] {};
 \draw [->] (a) to node[auto] {} (a1);
 \draw [->] (a) to node[auto] {} (a2);
 \draw [->] (a1) to node[auto] {} (a3);
  \draw [->] (a1) to node[auto] {} (a4);
    \draw [->] (a3) to node[auto] {} (a5);
        \draw [->] (a3) to node[auto] {} (a6);
 \end{scope}
   \begin{scope}[shift={(4,-3)}]
 \node (a) at ( 0,0) [place] {};
\node (a1) at ( -1/2,-1) [place] {};
\node (a2)  at ( 1/2,-1) [place] {};
\node (a3)  at ( -1,-2) [place] {};
\node (a4)  at ( 0,-2) [place] {};
\node (a5)  at ( -1/2,-3) [place] {};
\node (a6)  at ( 1/2,-3) [place] {};
 \draw [->] (a) to node[auto] {} (a1);
 \draw [->] (a) to node[auto] {} (a2);
 \draw [->] (a1) to node[auto] {} (a3);
  \draw [->] (a1) to node[auto] {} (a4);
    \draw [->] (a4) to node[auto] {} (a5);
        \draw [->] (a4) to node[auto] {} (a6);
 \end{scope}
  \begin{scope}[shift={(6,-3)}]
 \node (a) at ( 0,0) [place] {};
\node (a1) at ( -1/2,-1) [place] {};
\node (a2)  at ( 1/2,-1) [place] {};
\node (a3)  at ( 0,-2) [place] {};
\node (a4)  at ( 1/2,-2) [place] {};
\node (a5)  at ( 1,-2) [place] {};
\draw [->] (a) to node[auto] {} (a1);
\draw [->] (a) to node[auto] {} (a2);
\draw [->] (a2) to node[auto] {} (a3);
\draw [->] (a2) to node[auto] {} (a4);
\draw [->] (a2) to node[auto] {} (a5);
\end{scope}
  \begin{scope}[shift={(8,-3)}]
 \node (a) at ( 0,0) [place] {};
\node (a1) at ( -1/2,-1) [place] {};
\node (a2)  at ( 1/2,-1) [place] {};
\node (a3)  at ( 0,-2) [place] {};
\node (a4)  at ( 1,-2) [place] {};
\node (a5)  at ( 1/2,-3) [place] {};
\node (a6)  at ( 3/2,-3) [place] {};
 \draw [->] (a) to node[auto] {} (a1);
 \draw [->] (a) to node[auto] {} (a2);
 \draw [->] (a2) to node[auto] {} (a3);
  \draw [->] (a2) to node[auto] {} (a4);
    \draw [->] (a4) to node[auto] {} (a5);
        \draw [->] (a4) to node[auto] {} (a6);
 \end{scope}
   \begin{scope}[shift={(11,-3)}]
 \node (a) at ( 0,0) [place] {};
\node (a1) at ( -1/2,-1) [place] {};
\node (a2)  at ( 1/2,-1) [place] {};
\node (a3)  at ( 0,-2) [place] {};
\node (a4)  at ( 1,-2) [place] {};
\node (a5)  at ( -1/2,-3) [place] {};
\node (a6)  at ( 1/2,-3) [place] {};
 \draw [->] (a) to node[auto] {} (a1);
 \draw [->] (a) to node[auto] {} (a2);
 \draw [->] (a2) to node[auto] {} (a3);
  \draw [->] (a2) to node[auto] {} (a4);
    \draw [->] (a3) to node[auto] {} (a5);
        \draw [->] (a3) to node[auto] {} (a6);
 \end{scope}
\end{tikzpicture}
\end{center}
\caption{Schr\"oder trees contributing to the modular completion for $N=4, N_i=1$}
\label{fig:schroder-trees-4}
\end{figure}

For $N=4$, the modular completion shown in \cite[(F.10)]{Alexandrov:2019rth} involves contributions from the partitions $N=3+1=2+2=2+1+1=1+1+1+1$. The contributions from the 
2-body and 3-body partitions match our prescription exactly as in the rank 2 and rank 3 cases. The remaining partition with $N_i=4$ involves the contribution of the 11 Schr\"oder trees shown in \cref{fig:schroder-trees-4} (omitting Kronecker delta contributions): 
\be
\begin{split}
  \Delta \widehat{g}_{4, \mu} \, &= \, \frac{1}{8} \sum_{\substack{\scriptscriptstyle q_1+q_2+q_3+q_4 \\ =\mu+2}} \, q^{-\tfrac18 \; \sum_{i < j} \, (q_i - q_j)^2} \; y^{-9q_1-3q_2+3q_3+9q_4} \\
  & \times \Big\{ 
      E_{3}\!\Bigl(
        \tfrac{1}{\sqrt{3}},\;
        \tfrac{1}{\sqrt{6}},\;
        \tfrac{1}{\sqrt{2}};\;
        \sqrt{\tau_{2}}\,{\scriptstyle\bigl(-q_{1}+q_{2}+6\eta\bigr)},\;
        \sqrt{\tfrac{\tau_{2}}{3}}\,{\scriptstyle\bigl(2q_{3}-q_{1}-q_{2}+18\eta\bigr)},\;
        \sqrt{\tfrac{\tau_{2}}{6}}\,{\scriptstyle\bigl(-q_{1}-q_{2}-q_{3}+3q_{4}+36\eta\bigr)}
      \Bigr) \\
      &\quad + E_{2}\!\Bigl(
        \tfrac{1}{\sqrt{2}};\;
        \sqrt{\tfrac{\tau_{2}}{3}}\,{\scriptstyle\bigl(-q_{1}-q_{2}+2q_{3}+18\eta\bigr)},\;
        \sqrt{\tfrac{\tau_{2}}{6}}\,{\scriptstyle\bigl(-q_{1}-q_{2}-q_{3}+3q_{4}+36\eta\bigr)} 
      \Bigr) \; \sgn(q_1 - q_2) \\
      &\quad + E_{2}\!\Bigl(
        \tfrac{1}{2\sqrt{2}};\;
        \sqrt{\tfrac{\tau_{2}}{3}}\,{\scriptstyle\bigl(-2q_{1}+q_{2}+q_{3}+18\eta\bigr)},\;
        \sqrt{\tfrac{\tau_{2}}{6}}\,{\scriptstyle\bigl(-q_{1}-q_{2}-q_{3}+3q_{4}+36\eta\bigr)}
      \Bigr) \; \sgn(q_2 - q_3) \\
      &\quad + E_{2}\!\Bigl(
        \tfrac{1}{\sqrt{2}};\;
        \sqrt{\tau_{2}}\,{\scriptstyle\bigl(-q_{1}+q_{2}+6\eta\bigr)},\;
        \sqrt{\tfrac{\tau_{2}}{2}}\,{\scriptstyle\bigl(-q_{1}-q_{2}+q_{3}+q_{4}+24\eta\bigr)}
      \Bigr) \; \sgn(q_3 - q_4) \\
      &\quad +\,E_{1}\!\Bigl(
          \sqrt{\tfrac{\tau_2}{6}}\;{\scriptstyle\bigl(-q_{1}-q_{2}-q_{3}+3q_{4}+36\eta\bigr)}
        \Bigr) \, \sgn(q_2-q_1) \sgn(q_3-q_2) \\
      &\quad +\,E_{1}\!\Bigl(
          \sqrt{\tfrac{\tau_2}{6}}\;{\scriptstyle\bigl(-3q_{1}+q_{2}+q_{3}+q_{4}+36\eta\bigr)}
        \Bigr) \, \sgn(q_3-q_2) \sgn(q_4-q_3) \\
      &\quad +\,E_{1}\!\Bigl(
          \sqrt{\tfrac{\tau_2}{2}}\;{\scriptstyle\bigl(-q_{1}-q_{2}+q_{3}+q_{4}+24\eta\bigr)}
        \Bigr) \, \sgn(q_2-q_1) \sgn(q_4-q_3) \\
      &\quad - \, \sgn(q_2-q_1) \sgn(q_3-q_2) \sgn(q_4-q_3)
      \Big\}.
\end{split}
\ee
Disregarding $\eta$-shifts, this result can again be written in terms of a single maximum depth complementary error function,
\be
\begin{split}
  \Delta \widehat{g}_{4, \mu} \, &= \, \frac{1}{8} \sum_{\substack{\scriptscriptstyle q_1+q_2+q_3+q_4 \\ =\mu+2}} \, q^{-\tfrac18 \; \sum_{i < j} \, (q_i - q_j)^2} \; y^{-9q_1-3q_2+3q_3+9q_4} \\
  & \times M_{3} \Bigl(
        \tfrac{1}{\sqrt{3}},\;
        \tfrac{1}{\sqrt{6}},\;
        \tfrac{1}{\sqrt{2}};\;
        \sqrt{\tau_{2}}\,{\scriptstyle\bigl(-q_{1}+q_{2}+6\eta\bigr)},\;
        \sqrt{\tfrac{\tau_{2}}{3}}\,{\scriptstyle\bigl(2q_{3}-q_{1}-q_{2}+18\eta\bigr)},\;
        \sqrt{\tfrac{\tau_{2}}{6}}\,{\scriptstyle\bigl(-q_{1}-q_{2}-q_{3}+3q_{4}+36\eta\bigr)}
      \Bigr).
\end{split}
\ee
The observation that all complementary error functions but the one of maximal depth  cancel was the basis for \cite[Conjecture 6.1]{Alexandrov:2025sig}, where an exact formula was put forward the modular completion in one-modulus models (or more accurately, collinear D4-brane charges) at arbitrary rank, which also reproduces the Kronecker-delta terms. Mathematically, this implies that the modular completion can be written as an iterated Eichler integral, as was observed for $N=3$ in \cite{Manschot:2017xcr}.
Unfortunately we do not yet understand the physical origin of the Kronecker delta contributions. 

\subsection{Comparison to the instanton generating potential}
In view of the interpretation of the instanton generating potential (or its close cousin the contact potential) as the Witten index of the $\cN=2$ effective supergravity on $\IR^3$ \cite{Alexandrov:2014wca}, it is natural to ask if the localization computation in the supersymmetric quantum mechanics reproduces more generally the coefficient of the monomial
$\prod_i h_{p_i,\mu_i}$ in the expansion \eqref{eqn:defG}. In fact, up to the Gaussian factor 
$\Phi_1^{\int}$ which depends only on the total charge, 
the kernel of the theta series is identical to the one appearing in the modular completion \eqref{eqn:solRnr}, after replacing the moduli-independent vectors $\vb*{v}_k$ with their moduli-dependent counterpart $\vb*{u}_k$. We claim that the Witten index of the supersymmetric quantum mechanics correctly reproduces the sum over Schr\"oder trees in \eqref{eqn:Phitot}, even for non-collinear D4-brane charges. In particular, the contribution for the `purely moduli-dependent term' $\sign(c_1) \sign(c_1+c_2) \cdots
\sign(c_1+\cdots +c_{n-1})$ in the partial tree index reproduces the contribution $\widetilde{\cE}_{v_0}$ from the simplest Schr\"oder tree with a single node. To see this, it suffices to i) compare the Gram matrix of the vectors $\vb*{u}_{k}$ with respect to the quadratic form $\cQ$ defined below \eqref{eqn:def-Ev0-erf}, with the Gram matrix of the vectors $\sfH_{i\alpha}=1$ if $\alpha\geq i$, and zero otherwise, with respect to the quadratic form $\sfM$ defined in \eqref{eqn:defM}, and ii) compare the inner products $\vb*{u}_{k} \cdot \vb{x}$ with the products $\sfH^T c$.  
A straightforward computation shows that the Gram matrices agree, up to a simple rescaling,
\be
\vb*{u}_{k}^T \, \cQ \, \vb*{u}_{l} = 8 \, m_{\rm tot}^2 \, (\sfH^T \sfM \sfH)_{kl} =
8 \, m_{\rm tot} \qty(\sum_{i = 1}^{\min(k, l)} m_i) \qty(\sum_{i = \max(k, l)}^{n} m_i),
\ee
where $m_{\rm tot}=\sum_i m_i$,
while the inner products agree (up to a factor of $-m_{\rm tot}$), by construction,
\be
\vb*{u}_k \cdot \vb*{x} = - m_{\rm tot} \, (c_1 + \dots + c_k).
\ee
Thus, the error function $E_{n-1}$ of maximal depth is correctly reproduced, even for non-collinear D4-brane charges. Evidence that the supersymmetric quantum mechanics correctly reproduces the contributions from all Schr\"oder trees in \eqref{eqn:Phitot} comes from the matching of the modular completions in the previous subsection, although that evidence is restricted to the collinear case so far. We leave it as an open problem to prove agreement in general.

\section{Discussion}

In this note, we have computed the Witten index of the supersymmetric quantum mechanics of $n$ dyons for any $n\geq 2$ by localizing to time-independent configurations, following  \cite{Girardello:1983pw,Imbimbo:1983dg}.
The key observation is that the remaining integral over the positions of the constituents (and their fermionic partners) naturally splits into an integral over the phase space of $n$-centered supersymmetric ground states $\cM_n(\{\gamma_i,u_i\})$ at fixed values of the stability parameters 
$\{u_i\}$, with its standard Liouville measure $\omega^{n-1}$, times an integral over these parameters with a Gaussian weight centered around the physical FI parameters $\{c_i\}$. The integral over $\cM_n(\{\gamma_i,u_i\})$ is a locally constant function of the $u_i$'s, computable by localization with respect to rotations along a fixed axis \cite{Manschot:2011xc} or by applying the attractor flow tree decomposition \cite{Alexandrov:2018iao}. The resulting convolution with the Gaussian integral along the $u_i$'s then produces a linear superposition of generalized error functions $E_r$ of depth $r\leq n-1$, multiplied by $n-1-r$ sign functions of the FI parameters $c_i$ and DSZ products $\gamma_{ij}$. Upon expressing the $E_r$'s in terms of complementary error functions $M_{r'}$ with $r'\leq r$, we identified the terms proportional to $M_{r}$
with the contribution of the spectral asymmetry of the continuum of scattering states with $r+1$ constituents, which are in general themselves bound states of the fundamental constituents with charges $\gamma_i$. As an application, we reproduced the non-holomorphic terms appearing in the modular completion of the generating series of Vafa-Witten invariants on $\IP^2$, which count D4-D2-D0 bound states in the local CY geometry $K_{\IP^2}$, as well as the contributions to the so-called instanton generating potential. We expect that this agreement carries over to generating series of MSW indices in more general CY geometries, in particular arbitrary del Pezzo local surfaces, and gives a direct physical derivation of the
modular completion, which was previously derived through indirect arguments in the series of 
papers culminating with \cite{Alexandrov:2018lgp,Alexandrov:2019rth}. It would be interesting to see if one can also derive the modular completion of the generating series of single-centered indices \cite{Chattopadhyaya:2021rdi} by replacing the partial flow tree indices by the partial Coulomb indices \cite{Manschot:2013sya} counting collinear solutions.

While the agreement with the modular completion prescription holds to a remarkable level of detail, there are some intriguing discrepancies that indicate that our localization analysis may be too naive. In particular, in the two-body case, our result correctly captures the complementary error function $M_1$
in \eqref{eqn:Th2unref}, but misses the extra Gaussian term, which is crucial for modular invariance \cite{Manschot:2010sxc}. As understood in \cite{Alexandrov:2019rth}\footnote{The Gaussian term was reproduced in \cite{Murthy:2018bzs} by a localization computation in a gauged linear sigma model, where it also followed from an $\eta$-dependent shift in the refined index. It would be interesting to perform a similar computation in the quiver quantum mechanics reducing 
to our supersymmetric quantum mechanics on the Coulomb branch, along the lines of \cite{Hori:2014tda}.}, this term can be traced to an $\eta$-dependent shift in the refined index (where $\eta$ is proportional to $\Re\log y$), which is also absent in our  computation of the refined Witten index. In fact, our refined index computation also misses the contribution of the $A$-roof genus in the reduction to the symplectic leaves $\cM_n(\{\gamma_i,u_i\})$, which is necessary to produce the equivariant Dirac index which is supposed to count BPS states. Fortunately, under localization with respect to rotations, the $A$-roof genus only contributes powers of $(\log y)/(y-1/y)$ at isolated fixed points, powers which we have fixed by a suitable choice of normalization. These issues are likely related, and it would be desirable to develop a more rigorous treatment of localization in this system.

Another noteworthy issue is that our computation correctly reproduces the moduli-dependent
error functions appearing in the instanton generating potential (which plays the role of the Witten index for the 4D effective supergravity on $\IR^3$), but not the moduli-independent error functions appearing in the modular completion, which in general differ by a rescaling of their arguments (for example in the two-body case, a rescaling by 
 $\sqrt{ \frac{(p^3) (p_1 p_2 p)}{(p_1 p^2) (p_2 p^2)}}$). Fortunately, these rescalings  disappear when the D4-brane charge vectors are aligned, which is automatic in local CY geometries of the form $X=K_S$ where $S$ is a complex surface, or in compact CY geometries with $b_2(X)=1$. It would certainly be desirable to have a better understanding of this apparent discrepancy. 

Finally, we have not attempted to match the `exceptional' contributions to the modular completion which originate when the arguments of certain sign functions vanish. At such loci, the attractor point lies on a wall of marginal stability, and the dynamics is complicated by the existence of marginal bound states. It would be very interesting to have a physical derivation of these contributions, which are necessary to ensure a smooth unrefined limit.

\bigskip

\noindent {\bf Acknowledgments}: We are grateful to Sergey Alexandrov for useful discussions, and for comments on an early version of this note. We also thank Sameer Murthy for comments on the draft. This research is supported by the Agence Nationale de la Recherche under contract number ANR-21-CE31-0021. {\it For the purpose of Open Access, a CC-BY public copyright licence has been applied by the authors to the present document and will be applied to all subsequent versions up to the Author Accepted Manuscript arising from this submission. }

\bigskip

\appendix

\section{Proof of the key identity}\label{sec:proof-master-identity}

In this section, we prove the key identity \eqref{eqn:master-formula}, which we copy for the reader's convenience:
\be\label{eqn:master-formula-2}
  \prod_{i=1}^{n-1} \de^3{\va{x}_i} \det( \va{\nabla}_i{U_j} \otimes \va{\sigma}) = k_{n} \left(\prod_{i=1}^{n-1} \de{u_i} \right) \ \omega^{n-1},
\ee
where $\omega$ is the symplectic form defined in \eqref{eqn:defom} on the space $\cM_n(\{\gamma_i,u_i\})$ 
of multi-centered solutions with stability parameters $u_i= \sum_{j \neq i} \frac{\gamma_{i j}}{|\va{x}_i - \va{x}_j|}$ (we do not enforce the constraint $i<j$ in the sum, even though it is antisymmetric in $(ij)$), 
\be
\omega = \frac{1}{2} \sum_{i, j} \epsilon_{a b c} \nabla_i^a U_j \de{\mathrm{x}}_i^b \wedge \de{\mathrm{x}}_j^c.
\ee
In \eqref{eqn:master-formula-2}, $\va{\nabla}_i{U_j} \otimes \va{\sigma}$ is a $(2n-2) \times (2n-2)$ matrix with components $\sum_a \nabla^a_i U_j \qty(\sigma^a)_{\alpha \beta}$ with
$U_i = -\frac{1}{2} \qty(\sum_{j \neq i} \frac{\gamma_{i j}}{|\va{x}_i - \va{x}_j|} - c_i)$, 
and $k_n$ is a proportionality constant which will be fixed momentarily. 

Let us write $\va{x}_i = (x_i, y_i, z_i)$ and define $(n-1) \times (n-1)$ matrices $\Ux = \qty(\pdv{U_i}{x_j})$, $\Uy = \qty(\pdv{U_i}{y_j})$, $\Uz = \qty(\pdv{U_i}{z_j})$. The left hand side of \eqref{eqn:master-formula-2} can be written as the block matrix determinant
\be
  \det(\va{\nabla}{U} \otimes \va{\sigma}) = \det\mqty[\Uz & \Ux - i \Uy \\ \Ux + i \Uy & - \Uz] = (-1)^{n-1} \det(\Uz) \det(\mathsf{Q} + i \sfM),
\ee
where $\mathsf{Q} = \Uz + \Ux \Uz^{-1} \Ux + \Uy \Uz^{-1} \Uy$, and $\sfM = \Uy \Uz^{-1} \Ux - \Ux \Uz^{-1} \Uy$ are symmetric and anti-symmetric matrices respectively, such that $\mathsf{Q}+i\sfM$ is Hermitian. Here we have used the identity for the determinant of block matrices
\be
  \det\mqty[\mathsf{A} & \mathsf{B} \\ \mathsf{C} & \mathsf{D}] = \det(\mathsf{A}) \det(\mathsf{D} - \mathsf{C} \mathsf{A}^{-1} \mathsf{B}), \quad \text{if } \mathsf{A} \text{ is invertible}.
\ee
The right hand side of \eqref{eqn:master-formula-2} can be evaluated by noting that at constant $u_i(\mathbf{x}) = u_i$ (or equivalently constant $U_i = -\frac{1}{2}(u_i - c_i)$), $z_i$ can be expressed locally as a function of $x_j$ and $y_j$: $z_i = z_i(x, y; \, u)$. In other words, we can
use $(x_i,y_i)$ as local coordinates on the phase space $\cM_n(\{\gamma_i, u_i\})$. Thus, variations of $z$ can be traded for variations of $x$ and $y$,
\be\label{eqn:z-variation-constant-U}
  0 = \de{U_i} = \frac{\partial U_i}{\partial x_j} \de{x_j} + \frac{\partial U_i}{\partial y_j} \de{y_j} + \frac{\partial U_i}{\partial z_j} \de{z_j} \implies \frac{\partial z_i}{\partial x_j} = - \frac{\partial z_i}{\partial U_k} \frac{\partial U_k}{\partial x_j}, \quad \frac{\partial z_i}{\partial y_j} = - \frac{\partial z_i}{\partial U_k} \frac{\partial U_k}{\partial y_j},
\ee
or, as matrices, $
  \qty(\frac{\partial z_i}{\partial x_j}) = -\Uz^{-1} \Ux$, $ \qty(\frac{\partial z_i}{\partial y_j}) = -\Uz^{-1} \Uy$. We can now write the r.h.s. of \eqref{eqn:master-formula-2} as (here, summation convention is implied for $i, j = 1, \ldots, n-1$),
\be\label{eqn:computation-omega}
  \begin{aligned}
    \omega &= \frac{\partial U_i}{\partial z_j} \de{x_i} \wedge \de{y_j} + \frac{\partial U_i}{\partial x_j} \de{y_i} \wedge \de{z_j} + \frac{\partial U_i}{\partial y_j} \de{z_i} \wedge \de{x_j},
    \quad \ \de{z_i} = \frac{\partial z_i}{\partial x_k} \de{x_k} + \frac{\partial z_i}{\partial y_k} \de{y_k}, \\
    &= \frac{\partial U_i}{\partial z_j} \de{x_i} \wedge \de{y_j} + \qty(\frac{\partial U_i}{\partial x_j} \frac{\partial z_j}{\partial y_k} \de{y_i} \wedge \de{y_k} + \frac{\partial U_i}{\partial x_j} \frac{\partial z_j}{\partial x_k} \de{y_i} \wedge \de{x_k}) \\
    &\quad + \qty(\frac{\partial U_i}{\partial y_j} \frac{\partial z_i}{\partial x_r} \de{x_r} \wedge \de{x_j} + \frac{\partial U_i}{\partial y_j} \frac{\partial z_i}{\partial y_k} \de{y_k} \wedge \de{x_j}) \\
    &= \mathsf{A}_{i j} \, \de{x_i} \wedge \de{y_j} \, + \, \mathsf{B}_{i j} \, \de{x_i} \wedge \de{x_j} \, + \, \mathsf{C}_{i j} \, \de{y_i} \wedge \de{y_j},
  \end{aligned}
\ee
where
\be
  \mathsf{A}_{i j} = \frac{\partial U_i}{\partial z_j} - \frac{\partial U_j}{\partial x_l} \frac{\partial z_l}{\partial x_i} - \frac{\partial U_l}{\partial y_i} \frac{\partial z_l}{\partial y_j}, 
  \quad \mathsf{B}_{i j} = \frac{\partial U_l}{\partial y_j} \frac{\partial z_l}{\partial x_i},
  \quad \mathsf{C}_{i j} = \frac{\partial U_i}{\partial x_l} \frac{\partial z_l}{\partial y_j}.
\ee
Using \eqref{eqn:z-variation-constant-U}, and the symmetry $\nabla^a_i U_j = \nabla^a_j U_i$, etc, we get $\mathsf{A} = \mathsf{Q}$, and $\mathsf{B} = \mathsf{C} = -\Ux \Uz^{-1} \Uy$. From \eqref{eqn:computation-omega}, only the antisymmetric part of $\mathsf{B}$ and $\mathsf{C}$ are relevant, but $\mathsf{B} - \mathsf{B}^{T} = \mathsf{C} - \mathsf{C}^{T} = \Uy \Uz^{-1} \Ux - \Ux \Uz^{-1} \Uy = \sfM$. Thus we can write,
\be
    \omega = \frac{1}{2} \sum_{i, j, \alpha, \beta} \mathfrak{M}_{i \alpha, j \beta} \ \de{\xi^{\alpha}_i} \wedge \de{\xi^{\beta}_j}, \quad (\xi^1_i, \xi^2_i) = (x_i, y_i),
\ee
where $\mathfrak{M}$ is a $(2n-2) \times (2n-2)$ block matrix $\mathfrak{M} = \mqty[\sfM & \mathsf{Q} \\ -\mathsf{Q} & \sfM]$.
We can now compute the symplectic volume form as
\be
  \omega^{n-1} = 2^{n-1} (n-1)! \times \frac{1}{2^{n-1}} \ \mathrm{pf}\qty(\mathfrak{M}) \ \de{x_1} \wedge \de{y_1} \wedge \ldots \wedge \de{x_{n-1}} \wedge \de{y_{n-1}}.
\ee
Using \eqref{eqn:z-variation-constant-U}, and noting that $\de{U_i} = -\frac{1}{2} \de{u_i}$, we have
\be
  \de{u_1} \wedge \ldots \wedge \de{u_{n-1}} = 2^{n-1} \det(\Uz) \de{z_1} \wedge \ldots \wedge \de{z_{n-1}} + \ldots \ ,
\ee
where the dots include terms that have at least one factor of $\de{x_i}$ or $\de{y_i}$ which are annihilated when multiplied by $\omega^{n-1}$. Putting everything together, we obtain the r.h.s.
of \eqref{eqn:master-formula-2},
\be
  \left(\prod_{i=1}^{n-1} \de{u_i} \right) \ \omega^{n-1} = 2^{n-1} (n-1)! \ \det(\Uz) \ \mathrm{pf}\qty(\mathfrak{M}) \ \de^3{\mathbf{x}_1} \ldots \de^3{\mathbf{x}_{n-1}}.
\ee
The equality \eqref{eqn:master-formula-2} then follows from the identity for the Pfaffian of antisymmetric block matrices
\be
  \mathrm{pf}\mqty[\sfM & \mathsf{Q} \\ -\mathsf{Q} & \sfM] = \det(\mathsf{Q} + i \sfM).
\ee
provided the proportionality constant is set to 
\be
k_n = \frac{(-1)^{n-1}}{2^{n-1}(n-1)!}.
\ee


\providecommand{\href}[2]{#2}\begingroup\raggedright\endgroup


\begin{thebibliography}{10}

\bibitem{Strominger:1996sh}
A.~Strominger and C.~Vafa, \emph{Microscopic origin of the
  {B}ekenstein-{H}awking entropy}, {\emph{Phys. Lett.} {\bfseries B379} (1996)
  99} [\href{https://arxiv.org/abs/hep-th/9601029}{{\ttfamily
  hep-th/9601029}}].

\bibitem{Maldacena:1997de}
J.M.~Maldacena, A.~Strominger and E.~Witten, \emph{{B}lack hole entropy in
  {M}-theory}, {\emph{JHEP} {\bfseries 12} (1997) 002}
  [\href{https://arxiv.org/abs/hep-th/9711053}{{\ttfamily hep-th/9711053}}].

\bibitem{Alexandrov:2016tnf}
S.~Alexandrov, S.~Banerjee, J.~Manschot and B.~Pioline, \emph{{Multiple
  D3-instantons and mock modular forms I}},
  \href{https://doi.org/10.1007/s00220-016-2799-0}{\emph{Commun. Math. Phys.}
  {\bfseries 353} (2017) 379}
  [\href{https://arxiv.org/abs/1605.05945}{{\ttfamily 1605.05945}}].

\bibitem{Alexandrov:2017qhn}
S.~Alexandrov, S.~Banerjee, J.~Manschot and B.~Pioline, \emph{{Multiple
  D3-instantons and mock modular forms II}},
  \href{https://doi.org/10.1007/s00220-018-3114-z}{\emph{Commun. Math. Phys.}
  {\bfseries 359} (2018) 297}
  [\href{https://arxiv.org/abs/1702.05497}{{\ttfamily 1702.05497}}].

\bibitem{Alexandrov:2018lgp}
S.~Alexandrov and B.~Pioline, \emph{{Black holes and higher depth mock modular
  forms}}, \href{https://doi.org/10.1007/s00220-019-03609-y}{\emph{Commun.
  Math. Phys.} {\bfseries 374} (2019) 549}
  [\href{https://arxiv.org/abs/1808.08479}{{\ttfamily 1808.08479}}].

\bibitem{Alexandrov:2019rth}
S.~Alexandrov, J.~Manschot and B.~Pioline, \emph{{S-duality and refined BPS
  indices}}, \href{https://doi.org/10.1007/s00220-020-03854-6}{\emph{Commun.
  Math. Phys.} {\bfseries 380} (2020) 755}
  [\href{https://arxiv.org/abs/1910.03098}{{\ttfamily 1910.03098}}].

\bibitem{Alexandrov:2025sig}
S.~Alexandrov, \emph{{Mock Modularity at Work, or Black Holes in a Forest}},
  \href{https://doi.org/10.3390/e27070719}{\emph{Entropy} {\bfseries 27} (2025)
  719} [\href{https://arxiv.org/abs/2505.02572}{{\ttfamily 2505.02572}}].

\bibitem{Dabholkar:2012nd}
A.~Dabholkar, S.~Murthy and D.~Zagier, \emph{{Quantum Black Holes, Wall
  Crossing, and Mock Modular Forms}},
  \href{https://arxiv.org/abs/1208.4074}{{\ttfamily 1208.4074}}.

\bibitem{Zwegers-thesis}
S.~Zwegers, ``Mock theta functions.'' PhD dissertation, Utrecht University,
  2002.

\bibitem{MR2605321}
D.~Zagier, \emph{Ramanujan's mock theta functions and their applications (after
  {Z}wegers and {O}no-{B}ringmann)}, {\emph{Ast\'erisque} (2009) Exp. No. 986,
  vii}.

\bibitem{Vigneras:1977}
M.-F.~Vign\'eras, \emph{{S\'eries th\^eta des formes quadratiques
  ind\'efinies}}, {\emph{Springer Lecture Notes} {\bfseries 627} (1977) 227 }.

\bibitem{Pioline:2015wza}
B.~Pioline, \emph{{Wall-crossing made smooth}},
  \href{https://doi.org/10.1007/JHEP04(2015)092}{\emph{JHEP} {\bfseries 04}
  (2015) 092} [\href{https://arxiv.org/abs/1501.01643}{{\ttfamily
  1501.01643}}].

\bibitem{Murthy:2018bzs}
S.~Murthy and B.~Pioline, \emph{{Mock modularity from black hole scattering
  states}}, \href{https://doi.org/10.1007/JHEP12(2018)119}{\emph{JHEP}
  {\bfseries 12} (2018) 119}
  [\href{https://arxiv.org/abs/1808.05606}{{\ttfamily 1808.05606}}].

\bibitem{Troost:2010ud}
J.~Troost, \emph{{The non-compact elliptic genus: mock or modular}},
  \href{https://doi.org/10.1007/JHEP06(2010)104}{\emph{JHEP} {\bfseries 06}
  (2010) 104} [\href{https://arxiv.org/abs/1004.3649}{{\ttfamily 1004.3649}}].

\bibitem{Eguchi:2010cb}
T.~Eguchi and Y.~Sugawara, \emph{{Non-holomorphic Modular Forms and
  SL(2,R)/U(1) Superconformal Field Theory}},
  \href{https://doi.org/10.1007/JHEP03(2011)107}{\emph{JHEP} {\bfseries 03}
  (2011) 107} [\href{https://arxiv.org/abs/1012.5721}{{\ttfamily 1012.5721}}].

\bibitem{Ashok:2011cy}
S.K.~Ashok and J.~Troost, \emph{{A Twisted Non-compact Elliptic Genus}},
  \href{https://doi.org/10.1007/JHEP03(2011)067}{\emph{JHEP} {\bfseries 03}
  (2011) 067} [\href{https://arxiv.org/abs/1101.1059}{{\ttfamily 1101.1059}}].

\bibitem{Murthy:2013mya}
S.~Murthy, \emph{{A holomorphic anomaly in the elliptic genus}},
  \href{https://doi.org/10.1007/JHEP06(2014)165}{\emph{JHEP} {\bfseries 06}
  (2014) 165} [\href{https://arxiv.org/abs/1311.0918}{{\ttfamily 1311.0918}}].

\bibitem{Ashok:2013pya}
S.K.~Ashok, N.~Doroud and J.~Troost, \emph{{Localization and real Jacobi
  forms}}, \href{https://doi.org/10.1007/JHEP04(2014)119}{\emph{JHEP}
  {\bfseries 04} (2014) 119} [\href{https://arxiv.org/abs/1311.1110}{{\ttfamily
  1311.1110}}].

\bibitem{Harvey:2014nha}
J.A.~Harvey, S.~Lee and S.~Murthy, \emph{{Elliptic genera of ALE and ALF
  manifolds from gauged linear sigma models}},
  \href{https://doi.org/10.1007/JHEP02(2015)110}{\emph{JHEP} {\bfseries 02}
  (2015) 110} [\href{https://arxiv.org/abs/1406.6342}{{\ttfamily 1406.6342}}].

\bibitem{Gupta:2017bcp}
R.K.~Gupta and S.~Murthy, \emph{{Squashed toric sigma models and mock modular
  forms}},  \href{https://arxiv.org/abs/1705.00649}{{\ttfamily 1705.00649}}.

\bibitem{KumarGupta:2018rac}
R.~Kumar~Gupta, S.~Murthy and C.~Nazaroglu, \emph{{Squashed Toric Manifolds and
  Higher Depth Mock Modular Forms}},
  \href{https://doi.org/10.1007/JHEP02(2019)064}{\emph{JHEP} {\bfseries 02}
  (2019) 064} [\href{https://arxiv.org/abs/1808.00012}{{\ttfamily
  1808.00012}}].

\bibitem{Gaiotto:2019gef}
D.~Gaiotto and T.~Johnson-Freyd, \emph{{Mock modularity and a secondary
  elliptic genus}}, \href{https://doi.org/10.1007/JHEP08(2023)094}{\emph{JHEP}
  {\bfseries 08} (2023) 094}
  [\href{https://arxiv.org/abs/1904.05788}{{\ttfamily 1904.05788}}].

\bibitem{Alexandrov:2016enp}
S.~Alexandrov, S.~Banerjee, J.~Manschot and B.~Pioline, \emph{{Indefinite theta
  series and generalized error functions}}, {\emph{Selecta Mathematica}
  {\bfseries 24} (2018) 3927}
  [\href{https://arxiv.org/abs/1606.05495}{{\ttfamily 1606.05495}}].

\bibitem{Nazaroglu:2016lmr}
C.~Nazaroglu, \emph{{$r$-Tuple Error Functions and Indefinite Theta Series of
  Higher-Depth}},
  \href{https://doi.org/10.4310/CNTP.2018.v12.n3.a4}{\emph{Commun. Num. Theor.
  Phys.} {\bfseries 12} (2018) 581}
  [\href{https://arxiv.org/abs/1609.01224}{{\ttfamily 1609.01224}}].

\bibitem{Vafa:1994tf}
C.~Vafa and E.~Witten, \emph{{A Strong coupling test of S duality}},
  \href{https://doi.org/10.1016/0550-3213(94)90097-3}{\emph{Nucl.Phys.}
  {\bfseries B431} (1994) 3}
  [\href{https://arxiv.org/abs/hep-th/9408074}{{\ttfamily hep-th/9408074}}].

\bibitem{gottsche2018refined}
L.~G{\"o}ttsche and M.~Kool, \emph{{Refined {$SU(3)$} Vafa-Witten invariants
  and modularity}}, {\emph{Pure and Applied Mathematics Quarterly} {\bfseries
  14} (2019) 467} [\href{https://arxiv.org/abs/1808.03245}{{\ttfamily
  1808.03245}}].

\bibitem{Thomas:2018lvm}
R.P.~Thomas, \emph{{Equivariant $K$-Theory and Refined Vafa-Witten
  Invariants}}, \href{https://doi.org/10.1007/s00220-020-03821-1}{\emph{Commun.
  Math. Phys.} {\bfseries 378} (2020) 1451}
  [\href{https://arxiv.org/abs/1810.00078}{{\ttfamily 1810.00078}}].

\bibitem{Bringmann:2010sd}
K.~Bringmann and J.~Manschot, \emph{{From sheaves on $\mathbb{P}^2$ to a
  generalization of the Rademacher expansion}}, {\emph{Am. J. of Math.}
  {\bfseries 135} (2013) 1039}
  [\href{https://arxiv.org/abs/1006.0915}{{\ttfamily 1006.0915}}].

\bibitem{Manschot:2017xcr}
J.~Manschot, \emph{{Vafa-Witten Theory and Iterated Integrals of Modular
  Forms}}, \href{https://doi.org/10.1007/s00220-019-03389-5}{\emph{Commun.
  Math. Phys.} {\bfseries 371} (2019) 787}
  [\href{https://arxiv.org/abs/1709.10098}{{\ttfamily 1709.10098}}].

\bibitem{Alexandrov:2020bwg}
S.~Alexandrov, \emph{{Vafa\textendash{}Witten invariants from modular
  anomaly}}, \href{https://doi.org/10.4310/CNTP.2021.v15.n1.a4}{\emph{Commun.
  Num. Theor. Phys.} {\bfseries 15} (2021) 149}
  [\href{https://arxiv.org/abs/2005.03680}{{\ttfamily 2005.03680}}].

\bibitem{Alexandrov:2020dyy}
S.~Alexandrov, \emph{{Rank $N$ Vafa\textendash{}Witten invariants, modularity
  and blow-up}}, \href{https://doi.org/10.4310/ATMP.2021.v25.n2.a1}{\emph{Adv.
  Theor. Math. Phys.} {\bfseries 25} (2021) 275}
  [\href{https://arxiv.org/abs/2006.10074}{{\ttfamily 2006.10074}}].

\bibitem{Alexandrov:2024jnu}
S.~Alexandrov and K.~Bendriss, \emph{{Modular anomaly of BPS black holes}},
  \href{https://doi.org/10.1007/JHEP12(2024)180}{\emph{JHEP} {\bfseries 12}
  (2024) 180} [\href{https://arxiv.org/abs/2408.16819}{{\ttfamily
  2408.16819}}].

\bibitem{Alexandrov:2014wca}
S.~Alexandrov, G.W.~Moore, A.~Neitzke and B.~Pioline, \emph{{$\mathbb R^3$
  Index for Four-Dimensional $N=2$ Field Theories}},
  \href{https://doi.org/10.1103/PhysRevLett.114.121601}{\emph{Phys. Rev. Lett.}
  {\bfseries 114} (2015) 121601}
  [\href{https://arxiv.org/abs/1406.2360}{{\ttfamily 1406.2360}}].

\bibitem{Manschot:2010sxc}
J.~Manschot, \emph{{Stability and duality in N=2 supergravity}},
  \href{https://doi.org/10.1007/s00220-010-1104-x}{\emph{Commun. Math. Phys.}
  {\bfseries 299} (2010) 651}
  [\href{https://arxiv.org/abs/0906.1767}{{\ttfamily 0906.1767}}].

\bibitem{Denef:2002ru}
F.~Denef, \emph{{Quantum quivers and Hall/hole halos}}, {\emph{JHEP} {\bfseries
  10} (2002) 023} [\href{https://arxiv.org/abs/hep-th/0206072}{{\ttfamily
  hep-th/0206072}}].

\bibitem{Lee:2011ph}
S.~Lee and P.~Yi, \emph{{Framed BPS States, Moduli Dynamics, and
  Wall-Crossing}}, \href{https://doi.org/10.1007/JHEP04(2011)098}{\emph{JHEP}
  {\bfseries 04} (2011) 098} [\href{https://arxiv.org/abs/1102.1729}{{\ttfamily
  1102.1729}}].

\bibitem{Kim:2011sc}
H.~Kim, J.~Park, Z.~Wang and P.~Yi, \emph{{Ab Initio Wall-Crossing}},
  \href{https://doi.org/10.1007/JHEP09(2011)079}{\emph{JHEP} {\bfseries 09}
  (2011) 079} [\href{https://arxiv.org/abs/1107.0723}{{\ttfamily 1107.0723}}].

\bibitem{deBoer:2008zn}
J.~de~Boer, S.~El-Showk, I.~Messamah and D.~Van~den Bleeken, \emph{{Quantizing
  N=2 Multicenter Solutions}},
  \href{https://doi.org/10.1088/1126-6708/2009/05/002}{\emph{JHEP} {\bfseries
  05} (2009) 002} [\href{https://arxiv.org/abs/0807.4556}{{\ttfamily
  0807.4556}}].

\bibitem{Bena:2012hf}
I.~Bena, M.~Berkooz, J.~de~Boer, S.~El-Showk and D.~Van~den Bleeken,
  \emph{{Scaling BPS Solutions and pure-Higgs States}},
  \href{https://doi.org/10.1007/JHEP11(2012)171}{\emph{JHEP} {\bfseries 1211}
  (2012) 171} [\href{https://arxiv.org/abs/1205.5023}{{\ttfamily 1205.5023}}].

\bibitem{Descombes:2021egc}
P.~Descombes and B.~Pioline, \emph{{On the Existence of Scaling Multi-Centered
  Black Holes}},
  \href{https://doi.org/10.1007/s00023-022-01185-x}{\emph{Annales Henri
  Poincare} {\bfseries 23} (2022) 3633}
  [\href{https://arxiv.org/abs/2110.06652}{{\ttfamily 2110.06652}}].

\bibitem{Manschot:2010qz}
J.~Manschot, B.~Pioline and A.~Sen, \emph{{Wall Crossing from Boltzmann Black
  Hole Halos}}, \href{https://doi.org/10.1007/JHEP07(2011)059}{\emph{JHEP}
  {\bfseries 1107} (2011) 059}
  [\href{https://arxiv.org/abs/1011.1258}{{\ttfamily 1011.1258}}].

\bibitem{Manschot:2011xc}
J.~Manschot, B.~Pioline and A.~Sen, \emph{{A Fixed point formula for the index
  of multi-centered N=2 black holes}},
  \href{https://doi.org/10.1007/JHEP05(2011)057}{\emph{JHEP} {\bfseries 1105}
  (2011) 057} [\href{https://arxiv.org/abs/1103.1887}{{\ttfamily 1103.1887}}].

\bibitem{Alexandrov:2018iao}
S.~Alexandrov and B.~Pioline, \emph{{Attractor flow trees, BPS indices and
  quivers}}, \href{https://doi.org/10.4310/ATMP.2019.v23.n3.a2}{\emph{Adv.
  Theor. Math. Phys.} {\bfseries 23} (2019) 627}
  [\href{https://arxiv.org/abs/1804.06928}{{\ttfamily 1804.06928}}].

\bibitem{Girardello:1983pw}
L.~Girardello, C.~Imbimbo and S.~Mukhi, \emph{{On Constant Configurations and
  the Evaluation of the Witten Index}},
  \href{https://doi.org/10.1016/0370-2693(83)90224-1}{\emph{Phys. Lett. B}
  {\bfseries 132} (1983) 69}.

\bibitem{Imbimbo:1983dg}
C.~Imbimbo and S.~Mukhi, \emph{{Topological Invariance in Supersymmetric
  Theories With a Continuous Spectrum}},
  \href{https://doi.org/10.1016/0550-3213(84)90135-4}{\emph{Nucl. Phys. B}
  {\bfseries 242} (1984) 81}.

\bibitem{Beaujard:2020sgs}
G.~Beaujard, J.~Manschot and B.~Pioline, \emph{{Vafa\textendash{}Witten
  Invariants from Exceptional Collections}},
  \href{https://doi.org/10.1007/s00220-021-04074-2}{\emph{Commun. Math. Phys.}
  {\bfseries 385} (2021) 101}
  [\href{https://arxiv.org/abs/2004.14466}{{\ttfamily 2004.14466}}].

\bibitem{Dabholkar:2020fde}
A.~Dabholkar, P.~Putrov and E.~Witten, \emph{{Duality and Mock Modularity}},
  \href{https://doi.org/10.21468/SciPostPhys.9.5.072}{\emph{SciPost Phys.}
  {\bfseries 9} (2020) 072} [\href{https://arxiv.org/abs/2004.14387}{{\ttfamily
  2004.14387}}].

\bibitem{Gottsche:1990}
L.~G{\"o}ttsche, \emph{{The Betti numbers of the Hilbert scheme of points on a
  smooth projective surface}}, {\emph{Math.\ Ann.} {\bfseries 286} (1990) 193}.

\bibitem{Chattopadhyaya:2021rdi}
A.~Chattopadhyaya, J.~Manschot and S.~Mondal, \emph{{Scaling black holes and
  modularity}}, \href{https://doi.org/10.1007/JHEP03(2022)001}{\emph{JHEP}
  {\bfseries 03} (2022) 001}
  [\href{https://arxiv.org/abs/2110.05504}{{\ttfamily 2110.05504}}].

\bibitem{Manschot:2013sya}
J.~Manschot, B.~Pioline and A.~Sen, \emph{{On the Coulomb and Higgs branch
  formulae for multi-centered black holes and quiver invariants}},
  \href{https://doi.org/10.1007/JHEP05(2013)166}{\emph{JHEP} {\bfseries 05}
  (2013) 166} [\href{https://arxiv.org/abs/1302.5498}{{\ttfamily 1302.5498}}].

\bibitem{Hori:2014tda}
K.~Hori, H.~Kim and P.~Yi, \emph{{Witten Index and Wall Crossing}},
  \href{https://doi.org/10.1007/JHEP01(2015)124}{\emph{JHEP} {\bfseries 01}
  (2015) 124} [\href{https://arxiv.org/abs/1407.2567}{{\ttfamily 1407.2567}}].

\end{thebibliography}

\end{document}